\documentclass[]{aa}

\newcommand{\cd}{d$^{-1}$}
\newcommand{\kms}{km\,s$^{-1}$}
\newcommand{\ms}{m\,s$^{-1}$}

\newcommand{\vsini}{$v\sin{i}$}
\newcommand{\teff}{\ensuremath{T_{\rm eff}}}             % T_eff
\newcommand{\logg}{\ensuremath{\log g}}                     % log g

\usepackage{natbib}

\usepackage{graphicx}
\usepackage{txfonts}
\begin{document} 

  \title{Discovery of starspots on Vega
      \thanks{Based on observations obtained with the SOPHIE spectrograph at the 2m OHP telescope operated by the Institut National des Sciences de l'Univers (INSU) of the Centre National de la Recherche Scientifique of France (CNRS)}}

   \subtitle{First spectroscopic detection of surface structures on a normal A-type star.}

   \author{%
          T. B\"ohm\inst{\ref{inst:irap1},\ref{inst:irap2}} \and
          M. Holschneider\inst{\ref{inst:pots}} \and
          F. Ligni\`eres\inst{\ref{inst:irap1},\ref{inst:irap2}} \and
          P. Petit \inst{\ref{inst:irap1},\ref{inst:irap2}}\and 
          M. Rainer\inst{\ref{inst:brera}} \and
          F. Paletou\inst{\ref{inst:irap1},\ref{inst:irap2}} \and 
          G. Wade\inst{\ref{inst:royalmil}}\and
          E. Alecian\inst{\ref{inst:ipag}}\and
          H. Carfantan \inst{\ref{inst:irap1},\ref{inst:irap2}}\and
          A. Blaz\`ere\inst{\ref{inst:irap1},\ref{inst:irap2}}\and
          G.M. Mirouh\inst{\ref{inst:irap1},\ref{inst:irap2}}
           }

  % \offprints{T. B\"ohm}

   \institute{ 
              Universit\'e de Toulouse; UPS-OMP; IRAP; Toulouse, France\label{inst:irap1}\\
              \email{torsten.boehm@irap.omp.eu}
       \and       
              CNRS; IRAP; 14, avenue Edouard Belin, 31400 Toulouse, France\label{inst:irap2}
       \and
           Institut f\"ur Mathematik, Universit\"at Potsdam, DYCOS, 14469 Potsdam, Germany\label{inst:pots}
       \and
 	    INAF - Osservatorio Astronomico di Brera, via E. Bianchi 46, 23807 Merate, Italy\label{inst:brera}     
       \and       
           Observatoire de Grenoble, Universit\'e Joseph Fourier, IPAG, 38041 Grenoble, France\label{inst:ipag}
       \and
           Dept. of Physics, Royal Military College of Canada, PO Box 17000, Stn Forces, Kingston, Canada K7KK 7B4\label{inst:royalmil}
     }
 \date{Received November $27^{th}$, 2014; accepted march 23$^{rd}$, 2015}

% \abstract{}{}{}{}{} 
% 5 {} token are mandatory
 
\abstract
  % context heading (optional)
  % {} leave it empty if necessary  
{The theoretically studied impact of rapid rotation on stellar evolution needs to be confronted with the results of high resolution spectroscopy-velocimetry observations. Early type stars present a perfect laboratory for these studies. The prototype A0 star Vega has been extensively monitored in the last years in spectropolarimetry. A weak surface magnetic field has been detected, potentially leading to a (yet undetected) structured surface.}
    % aims heading (mandatory)
   {The goal of this article is to present a thorough analysis of the line profile variations and associated estimators in the early-type standard star Vega (A0) in order reveal potential activity tracers, exoplanet companions and stellar oscillations.}
  % methods heading (mandatory)
   {Vega was monitored in high-resolution spectroscopy with the velocimeter Sophie/OHP. A total of 2588 high S/N spectra was obtained during 5 nights (August 2012) at  R = 75000 and covering the visible domain. For each reduced spectrum,  Least Square Deconvolved (LSD) equivalent photospheric profiles were calculated with a \teff = 9500 and \logg = 4.0 spectral line mask. Several methods  were applied to study the dynamic behavior of the profile variations (evolution of radial velocity, bisectors, vspan, 2D profiles, amongst others).}
  % results heading (mandatory)
 {We present the discovery of a starspotted stellar surface in an A-type standard star with faint spot amplitudes $\Delta {\rm F/Fc} \sim 5 \times 10^{-4}$. A rotational modulation of spectral lines with a period of rotation P = 0.68\,d has clearly been exhibited, confirming the results of previous spectropolarimetric studies. Either a very thin convective layer can be responsible for magnetic field generation at small amplitudes, or a new mechanism has to be invoked in order to explain the existence of activity tracing starspots.}
  % conclusions heading (optional), leave it empty if necessary 
   {This first strong evidence that standard A-type stars can show surface structures opens a new field of research and asks the question about a potential link with the recently discovered weak magnetic field discoveries in this category of stars.}
   
   
 
\maketitle

\section{Introduction}
\label{intro}

The role rapid rotation plays on the stellar interior and its evolution represents a very challenging research topic as of today. Rapidly rotating stars reveal many unanswered questions in the domain of observations, theory and modeling. The only known way  to study stellar interiors is through asteroseismology and its associated observational techniques, which are either based on photometry or high-resolution spectroscopy.  In addition, the detection and observation of activity tracing structured stellar surfaces can also contribute significant constraints on stellar evolution models.

The recent detection of a very weak magnetic field in Vega was reported in \cite{lignieres2009}. The -0.6$\pm$0.3\,G the disk-averaged line-of-sight component of the surface magnetic field can reach peak values of 7\,G \citep{petit2014}. \cite{lignieres2009} and later \cite{petit2010} concluded on the fact that  a previously unknown type of magnetic stars exist in the intermediate-mass domain, and that Vega may well be the first confirmed member of a much larger, as yet unexplored, class of weakly-magnetic stars. The Zeeman-Doppler imaging of the magnetic field topology \citep{petit2010,petit2014} showed that apart from a prominent polar magnetic region a few other magnetic spots are reconstructed at lower latitude. \cite{petit2010} conclude that an important help for distinguishing between a potential fossil or dynamo origin of the magnetic field would be to investigate the long-term stability of the observed field geometry, as a dynamo-generated field is likely to show some temporal variability. Detecting surface structures in unpolarized light would largely contribute to a better understanding of the origin of magnetic fields in these stars, and the role the rotation could play. 

Large scale surveys of A-type stars with the Kepler satellite revealed low frequency periodic variations in 28\% of the sample interpreted as linked to the stellar rotation frequency and associated rotational modulation (\cite{balona2011}). The author suggests that the light variations in A-type stars may possibly be due to starspots or other corotating structures and that A-star atmospheres may not be quiescent as previously supposed. In a more recent article on Kepler data, \cite{balona2013} reports that in 875 A-type stars (40\%), photometric indications for the period of rotation are detected. From the amplitude distribution he concludes that the sizes of starspots in A-type stars are similar to the largest sunspots. He concludes that A-type stars are active and, like cooler stars, have starspots and flares.
Providing direct detection of starspotted surfaces in A-type stars is therefore a necessary next step in order to ascertain this thesis.

The A0 photometric standard star Vega is a rapid rotator. Vega is seen pole-on with an inclination angle of i=$5-7^{\circ}$, a result ascertained by photometric \citep{gray1985}, spectroscopic \citep{gulliver1994} and interferometric \citep{aufdenberg2006, peterson2006, monnier2012} observations. Vega's \vsini \, has been determined by \cite{hill2004}:  21.9$\pm$ 0.1\,\kms. A new determination by \cite{yoon2010} concludes, model-dependend, on slightly lower values: 20.48$\pm$ 0.11\,\kms or 20.80$\pm$ 0.11\,\kms.

While spectroscopic analysis and modeling by \cite{takeda2008} and \cite{hill2010} slightly differ in equatorial velocities  ($\sim$ 175 km\,s$^{-1}$ versus 211$\pm$ 4\,\kms, respectively), first interferometric studies by \cite{peterson2006} and \cite{aufdenberg2006} conclude on a much higher equatorial velocity of $\sim$ 275 km\,s$^{-1}$. 

\cite{petit2010} announced that the short-term evolution of polarized signatures in Vega is consistent with a rotational period of 0.732 $\pm$ 0.008 d, the error bar being underestimated (private com.). \cite{alina2012} worked on the NARVAL/TBL 2010 data set   \citep{alina2012, boehm2012} and detected a period of 0.678 $\pm^{0.036}_{0.029}$ d. \cite{Budkovskaya2013} analyzed the results of 1312 longitudinal magnetic field measurements obtained during 15 years of observations performed at the Crimean Astrophysical Observatory and proposed that the magnetic field variations are caused by stellar rotation; their derived stellar rotation period corresponds to 0.6225503 days. 
Both determined periods are significantly longer than values expected from earlier interferometry-based models, but compatible with the results of spectroscopic modeling of gravity darkened photospheric lines in Vega \citep{takeda2008}. Based on new interferometric CHARA/MIRC observations, \cite{monnier2012} concluded that a more slowly rotating model was compatible with the new interferometric data, thereby reconciling these new results with those determined by spectroscopy. 

As reviewed by \cite{gray2007}, Vega was used since more than 150 years as a photometric and spectrophotometric standard. Still, very low photometric variability was occasionally reported at the 1-2\% level.
\cite{hill2010} report low level variations  of Ti II 4529\,AA\, profiles on the time basis of several years. First evidence of pulsations in Vega was suspected in three data sets (3 nights NARVAL/TBL 2008 - 3 nights ESPADONS/CFHT 2009 - 5 nights TBL 2010) corresponding to a total of 4478 quasi-continuous high-resolution (R $>$ 65000) echelle spectra \cite{boehm2012}. Least square deconvolved (LSD) profiles \citep{donati1997} were obtained for each spectrum representing the photospheric absorption profile potentially deformed by the presence of pulsations, and telluric lines were used as a velocity reference.  All three data sets revealed the presence of residual periodic variations with the following frequencies and amplitudes: 5.32 and 9.19\,\cd\, (A $\approx$ 6\ms) in 2008, 12.71 and 13.25\,\cd\, (A $\approx$ 8\ms) in  2009 and 5.42 and 10.82\,\cd\, (A $\approx$ 3-4\ms) in 2010. 
However, due to a lack of intrinsic stability of the employed spectropolarimeters it was too early to conclude that the variations were due to stellar pulsations, and it was concluded that their confirmation with a highly stable spectrograph was the necessary next step. The results of a 5 night survey with the highly stabilized spectrograph SOPHIE/OHP is presented in this article.

Section \ref{obs} describes the observations and data reduction, Section \ref{results} presents the results concerning rotation and surface spots, potential exoplanet companions and stellar oscillations and provides the discussion of the results, Section \ref{disc} presents the conclusion of the article.

\section{Observations and data reduction}
\label{obs}

\begin{table*}
\caption[]{Log of the spectroscopic observations of Vega. (1) Date  of the observation (UT). (2)  and (3) Barycentric Julian date (mean observation, 2,450,000+) of the first and the last stellar spectrum of the night, respectively; (4) Number of high resolution Vega spectra obtained; (5) exposure time (sec); (6) total hours covered on the sky; (7) nightly average and standard deviation of signal to noise ratio per resolved element at 520\,nm.}
\label{table:log}
\centering
\begin{tabular}{ccccccl}
\hline\hline
 Date	& BJD$_{\rm first}$	& BJD$_{\rm last}$	& N$_{\rm spec}$	& t$_{\rm exp}$ (sec)& t$_{\rm cov}$ (hrs) & S/N \\
(1)		&  (2)  			&  (3)        			&   (4)         		&        (5)                       &     (6)                        &   (7)  \\
\hline
Aug. 2 2012 &  6142.3308	&	6142.6238 	& 425	                   &	17-30		&   7.0     			& 884$\pm$202\\ 
Aug. 3 2012 &  6143.3528       &        6143.6436         & 629                           &	10-15		&   7.0   			& 925$\pm$145\\
Aug. 4 2012 &  6144.3412       &        6144.6423        &  628                          &	10-17		&   7.2    			& 850$\pm$121\\ 
Aug. 5 2012 &  6145.3788	&      6145.6442         & 402                           &	13-17		&   6.4   			& 766$\pm$103\\
Aug. 6 2012 &  6146.3444       &        6146.6423        & 504                           &	17-25	 	&   7.1     			& 808$\pm$115\\ \hline
\end{tabular}
\end{table*}

In August 2012, 2588 high-resolution ($R = 75000$) spectra of Vega were acquired during 34.7 hrs on 5 consecutive nights, using the ultrastable echelle spectrograph SOPHIE/OHP. The log of observations is reported in Tab. \ref{table:log}.

The typical S/N per resolved element was around 800 at 520\,nm. Spectra with low signal to noise (S/N $<$ 300 per resolved element) were rejected. Due to the excellent intrinsic stability of the spectrograph (of the order of 2-3\ms using simultaneous Th/Ar calibration) no long-term trends related to instrumental shifts were present in the data set. The data reduction procedure was reduced to its minimum based on the standard SOPHIE data reduction pipeline, in order to avoid the introduction of any artifacts.

We then calculated for all stellar spectra photospheric LSD-profiles \citep{donati1997} using the original well-established prescription and the original code. The resulting equivalent photospheric profiles have been used in many domains, such as magnetic structures (in polarimetry), ZDI (Zeeman-Doppler Imaging), asteroseismology, Doppler imaging etc. The article by  \cite{reiners2003} demonstrates that LSD does not modify the original shape of line profiles. Recently, \cite{kochukhov2010} have analyzed in depth the behaviour of LSD, mostly concerning the polarimetric aspects. LSD applied on intensity spectra corresponds to a (weighted) cross-correlation technique. This implies that all conclusions obtained concerning the validity of cross-correlation techniques applied to Doppler searches for exoplanets do apply also for LSD. This includes radial velocities, but also bisectors. The regime in which LSD is used in this article concerns fluctuations of several m/s, which implies that our study is located very comfortably within the application regime of cross-correlation techniques, as demonstrated by many publications on exoplanets (see for example the capacity of these methods on articles concerning HARPS observations on alpha Cen by Dumusque et al. \citep{dumusque2012} 
or the article by Bazot et al. \citep{bazot2011} on 18 Sco. In polarimetry, LSD has proved to be extremely sensitive to signal variations of the order of 10$^{-5}$  of the continuum \citep{auriere2009, auriere2010,  lignieres2009}. LSD conserves the asymmetries of the spectral lines and therefore also of the bisectors, as demonstrated by articles on cool stars (e.g. articles on $\xi$ Boo: \cite{petit2005, morgenthaler2012}. In summary, LSD profiles have been used successfully  to study line profile variations in many asteroseismological articles (eg. \cite{uytterhoeven2008, nardetto2014, mantegazza2012, antonello2006}). This shows that the LSD technique is indeed suited to study small line asymmetries.

To calculate the LSD profiles, we used a mask based on a Kurucz atmospheric model of spectral type A0, well matching the fundamental parameters of Vega: \teff = 9500\,K,  \logg = 4.0 and solar metallicity (it should be kept in mind that Vega is slightly underabundant). The adopted effective temperature lies in between the polar and equatorial temperatures as published in \cite{takeda2008} and \cite{hill2010}, the latter concluding on a mean temperature of 9560\,$\pm$ 30\,K. As shown by \cite{takeda2008}, due to important gravitational darkening in this rapidly rotating star, weak lines show flat-bottomed profiles, very different from the well-known rotationally broadened profiles observed in the stronger lines (see also Figs. 1 and 2 of \cite{boehm2012}). In order to avoid mixing of very different photospheric line profile shapes, we selected only the stronger lines with an unbroadend depth between 0.3 and 1.0 and excluded hydrogen Balmer lines. We kept a subset of 299 lines covering the visible acquisition domain of SOPHIE/OHP from 389 - 627\,nm.  All lines with a photospheric shape were kept, without any selection of particular line parameters. A detailed analysis of the results related to different line lists reveals that there is some small impact of the selected rejection threshold (0.1, 0.2, 0.3 ...) on the detailed results of the different analysis presented in this study. However, the overall conclusions remain unaffected.

In a next step, we derived for each LSD profile several quantities. The bisector of the line was determined by calculating for each depth the exact position of the left and right side of the profile (via interpolation of the local profile), the bisector being the average of these two positions as a function of profile depth. Vspan was calculated for each profile, measuring the difference of the upper and lower part of the bisector (a kind of skewness); for this, we worked in relative profile height, the bottom of the profile set at 0., the continuum at 1. Vspan was calculated as the difference of the medians of upper [0.35, 0.5] and lower [0.1, 0.25] bisector ranges. Ranges were defined checking out different combinations in order to produce the most clear periodograms.  Optimal ranges are lying rather low in the profile, which can be understood having in mind the nearly pole-on position of the star, since equatorial contributions are barely seen (those which cross the full line profile from -\vsini\, to +\vsini).

\begin{figure}
 \includegraphics[width=9cm]{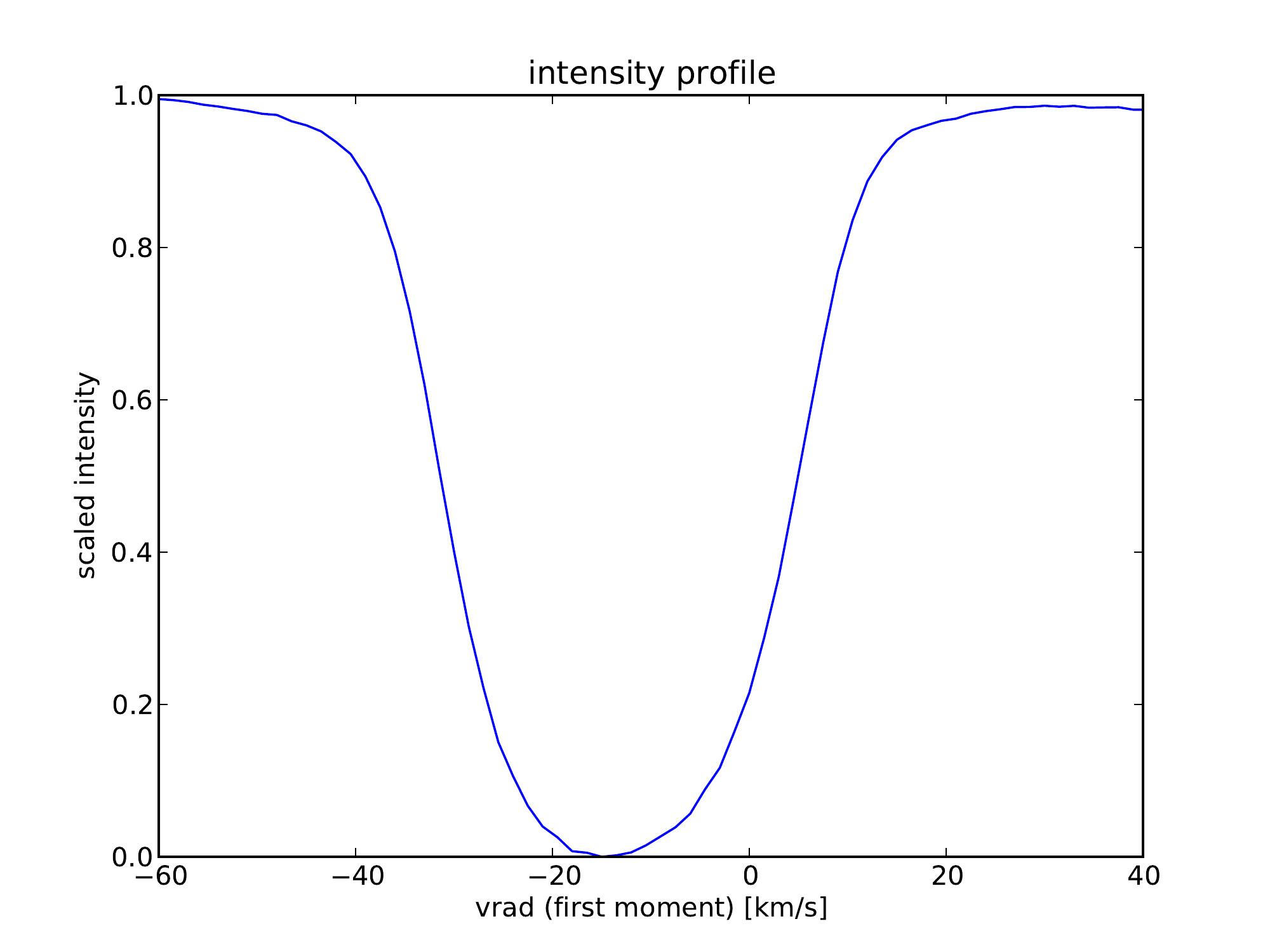}
 \caption{Mean equivalent photospheric (LSD) profile of Vega, stronger lines only, rescaled in depth.}
 \label{fig:intens}
\end{figure}

\begin{figure}
\includegraphics[width=11.5cm]{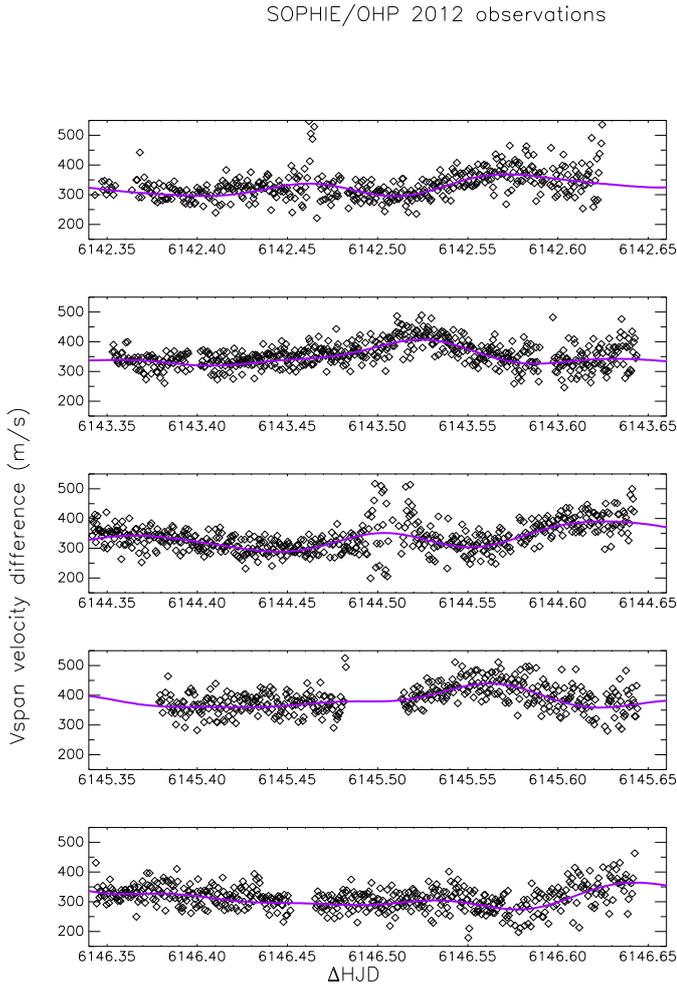}
\caption{Vspan variations of Vega (SOPHIE/OHP) 2012. Each frame represents one night's observations. Superimposed
(continuous line) is the result of the corresponding frequency analysis  (table \ref{table:freqvspan}).
Time is expressed in BJD = 2450000 + $\Delta$\,BJD. }
 \label{fig:time}
 \end{figure}

\begin{figure*}
 \begin{center}
 \includegraphics[width=16cm]{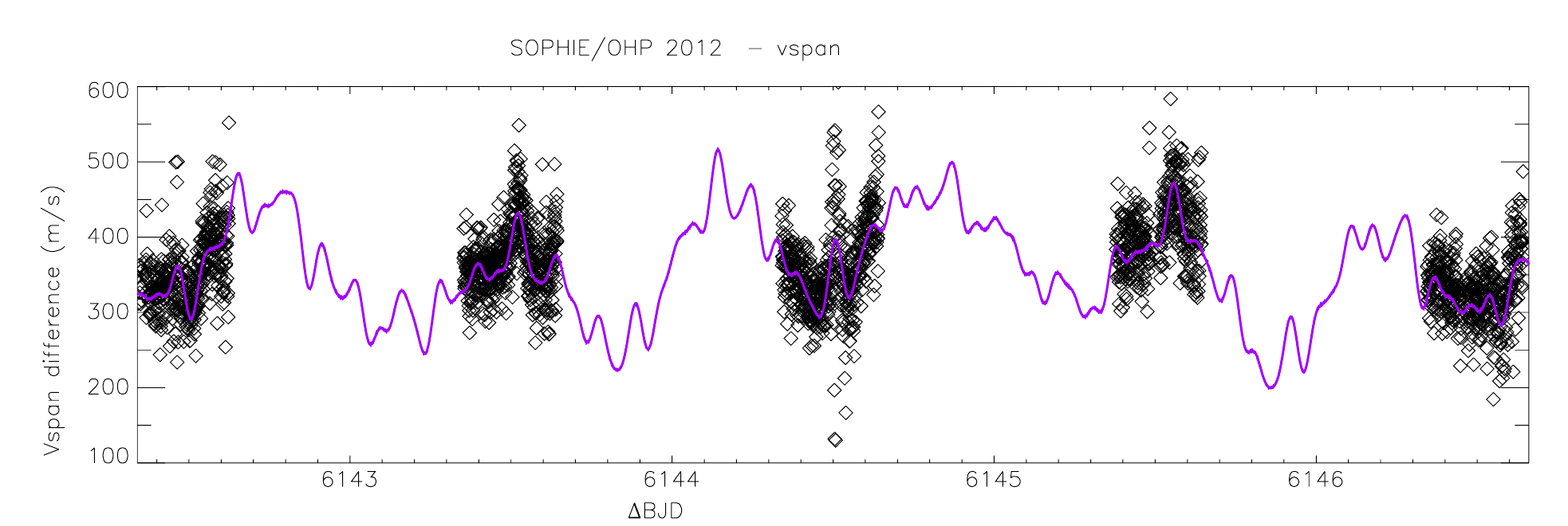}\\
   \caption{Low frequency vspan variations of Vega (SOPHIE/OHP) 2012. Superimposed
(continuous line) is the result of the corresponding frequency analysis as listed  in Tables \ref{table:freqvspan}.
Time is expressed in BJD = 2450000 + $\Delta$\,BJD. The period of rotation can clearly be seen in this figure.} 
   \label{fig:time_vspan_tot}
    \end{center}
 \end{figure*}

Fig. \ref{fig:intens} shows the corresponding mean rescaled LSD profile. The variation of vspan with time is shown in Figs.  \ref{fig:time} and \ref{fig:time_vspan_tot}, the latter one showing the low frequency variations.

Radial velocity (RV) can be determined in several ways: cross-correlation with an average profile, fitting of gaussians or rotationally broadened profiles, first moment determination and median determination of the lower part of the bisector, amongst others. Having computed the different methods on our data set, we present in this study the radial velocities determined by  i) the first moment of the profile and ii) the median of the lower part of the line bisector range [0.15, 0.3], the bottom of the line being particularly sensitive to potential spot signatures crossing the line profile. The first moment has been calculated in the following way:

Let $I_n$, $n=1,\dots N$, be a collection of intensities within a velocity bin with center velocity $v_n$. 
The continuum level is at $I=1$. Since we consider absorption lines, we have $0\leq I_n \leq 1$. 
The radial velocity estimate is then defined through
$$
\hat v_{r} = \frac{\sum_{n=1}^N (1-I_n)\, v_n}{ \sum_{n=1}^N (1-I_n)}.
$$
In practice the bins may be preselected based on some threshold $c$ such that $I_n \leq c$. In our case we used c = 0.5 in order to be sensitive to the most variable parts of the LSD profile. The depth c has some effect on the radial velocity determination, but the conclusions of this article remain valid as a whole. 

As can be seen in Tables \ref{table:freqbis} and \ref{table:freqvmean}, for a given frequency, the amplitude of the radial velocity variation strongly varies between the two estimators (first moment, bisector). This fact is not really surprising taking into
account that both determinations are significantly different, and that a 10\,\ms  \,value corresponds to only 2 $\times$ 10$^{-4}$ of the overall width of the profile ($\simeq$ 2\, \vsini, i.e. 44\,\kms).

In a next step, we worked on the line profile variations themselves (as a function of velocity). The very small variations we discovered never exceeded 10$^{-3}$ of the continuum. In traditional spectroscopic data sets such small variations are in generally not seen, mostly due to insufficiently precise data reduction or continuum normalization, amongst other issues. In our data set, continuum normalisation is performed several times. The overall spectra are normalized following reduction of the data, then the LSD procedure contains a local renormalisation, and eventually very tiny residuals (profile-to profile variations) in the continuum level are renormalized in the resulting LSD profiles. Clearly, the precision of normalization cannot be responsible for any of the observed variability.
At these tiny levels of variation a particular care has to be taken in order to avoid misinterpretation due to residual instrumental effects (see e.g. Sect. \ref{spot}). 
 
\section{Results and discussion}
\label{results}

\subsection{Periodicity analysis of vspan and radial velocity}
\label{radial}

We started the search for periodicities in our line profile estimators by analyzing the vspan data, which are insensitive to radial velocity calibration data (since they measure
profile asymmetries to some extent, and no shifts). This asymmetry measurement reveals to be robust and provides similar frequency values, independently of the precise depth boundaries chosen. 

\begin{figure*}
 \begin{center}
\includegraphics[width=13cm]{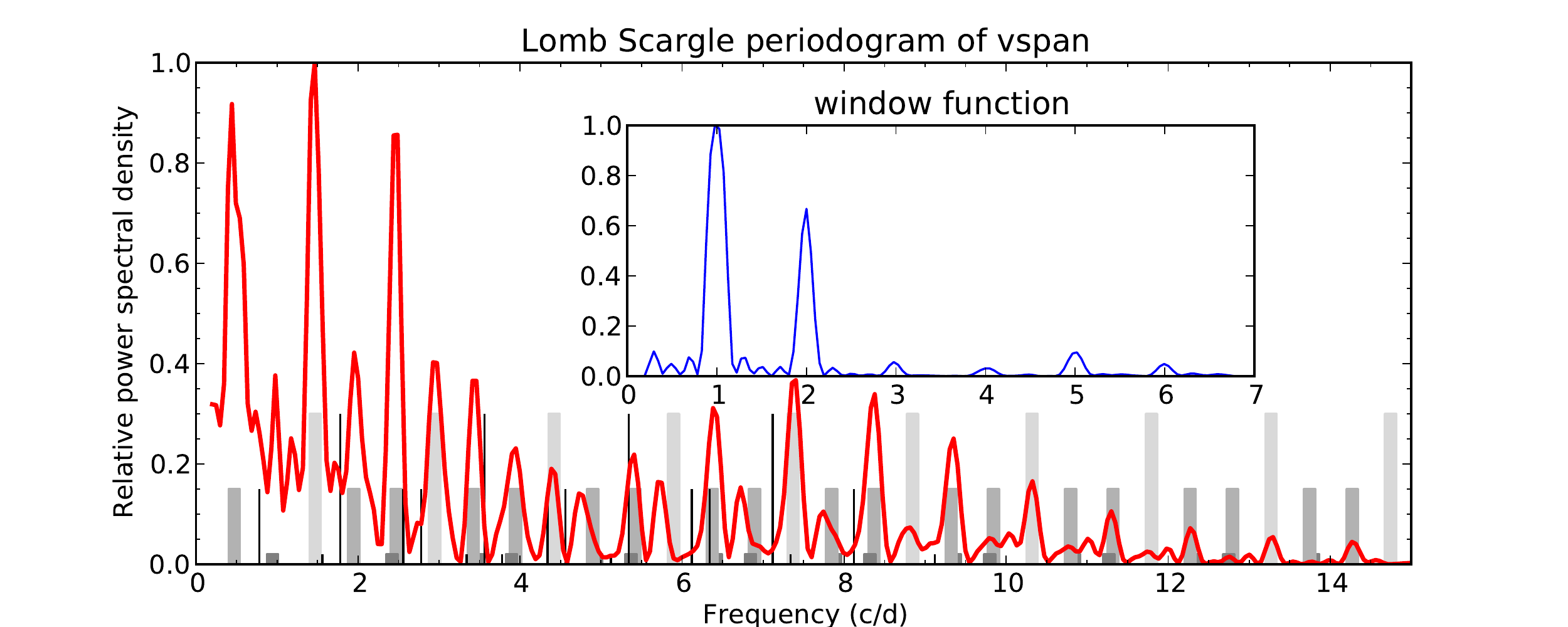}\\
\includegraphics [width=13cm]{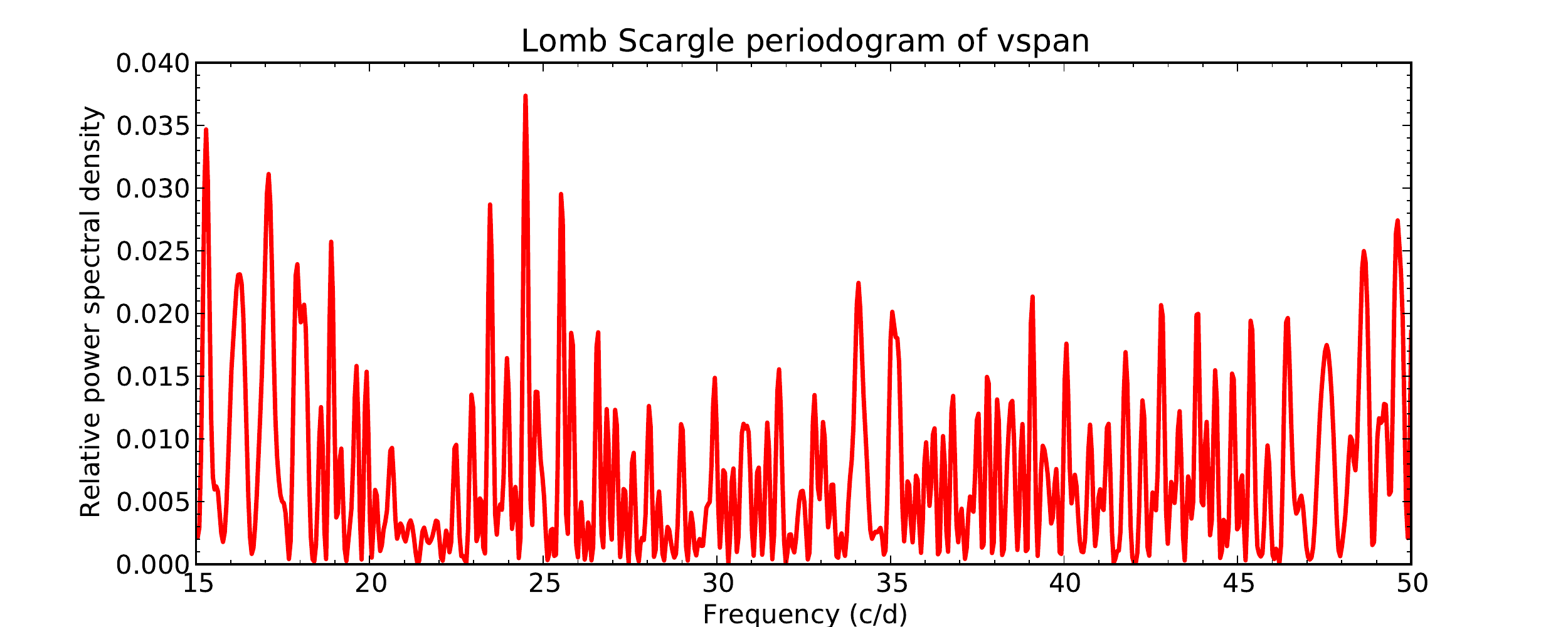}
   \caption{Lomb Scargle periodogram of vspan, including the window function of the data set. 
   The grey bars at 0.3 height indicate the rotational frequency of the star as determined by \cite{alina2012}, ie. 1.47 \cd, as well as its harmonics (multiple of this frequency), its width corresponds to the error bar. The lower bars at
  0.15 and 0.02 height indicate the position of the $\pm$ 1 and 2 day aliases with respect to the rotational frequency comb, respectively. The thin vertical black line indicates the same kind of frequency grid for a frequency of 1.77 \cd (see Sect. \ref{osc}). }
   \label{fig:ls_spec_vspan}
    \end{center}
 \end{figure*}

The Lomb Scargle periodogram of vspan is shown in Fig.    \ref{fig:ls_spec_vspan}. In a recent paper, \cite{alina2012} had determined for Vega a P$_{\rm rot} $= 0.678 $\pm^{0.036}_{0.029}$\,d based on spectropolarimetric observations. We overplotted in grey bars at 0.3 height the corresponding rotational frequency of  1.47 \cd, as well as its harmonics (multiple of this frequency), the width of the bars corresponding to the reported error bars. The lower bars at 0.15 and 0.02 height indicate the position of the 
$\pm$ 1 and 2 day aliases with respect to the rotational frequency comb, respectively.

An important result is the fact that many detected periodicities in the range 0 - 15 \cd \,correspond to the stellar rotation frequency (and its harmonics and aliases),  and are in perfect agreement with the rotation period as published by  \cite{alina2012}.  This provided us with a first evidence that surface structures should cross the photospheric equivalent line profile, leading to a rotational modulation of
the vspan parameter.  

Table \ref{table:freqvspan} present the corresponding periodicity analysis of this vspan time series based on the SigSpec tool \citep{reegen2007}. SigSpec calculates the spectral significance  of an amplitude A by: $sig (A) = - \log [\Phi_{\rm FA} (A)]$,  where $\Phi_{\rm FA}$ indicates the false alarm probability. To do this, SigSpec computes the probability density function (pdf) of white noise in the Fourier space. The integration of the pdf yields the false-alarm probability that white noise in the time domain produces an amplitude of at least A. The sig threshold for the determination of the prewhitening sequence was equal to 4, which indicates that the considered amplitude level is due to noise in less than one out of 10$^4$ cases. Only such frequencies have been listed in tables \ref{table:freqvspan}, \ref{table:freqbis} and \ref{table:freqvmean}. 
These frequencies are formally reliable, however, one must keep in mind that the measured entities can carry intrinsic errors and biases, which cannot easily be discarded before the subsequent frequency analysis.

\cite{kurtz1999} estimate the uncertainty in the frequency to be approximately 1/(4$\Delta T$), where $\Delta T$ is the time span of the data set. In our case this would yield an uncertainty of 0.05\,d$^{-1}$. The last column of the table proposes some identification with the rotation comb (the star's rotation frequency F$_{\rm rot}$, its harmonics, and $\pm$1d window function aliases only). For each frequency combination we assumed that we correctly identified them, and calculated the distance to the expected position. The average distance is only 0.043\cd and tends therefore to confirm our identification.

\begin{table}[!ht]
\caption{Frequencies and amplitudes of vspan variations measured on the LSD-profiles, and possible identifications. Only frequencies with a significance value above 4 are presented. }
\label{table:freqvspan}
\centering
\begin{tabular}{|c|c|c|c|c|c|} \hline
ID      & freq.    &   $A$       & Comment\\
       	 & \cd\,     &   \ms        &          \\ \hline
F1 	 &   1.457  & 69.32 	   &        F$_{\rm rot}$  \\
F2     &    2.02  & 22.93         &    2 F$_{\rm rot}$-1.  \\
F3     &    0.40  &  47.52        &       \\
F4     &    8.34 & 13.30          & 5 F$_{\rm rot}$ + 1. \\
F5     &    11.31 &  14.17         & 7 F$_{\rm rot}$ + 1. \\
F6     &    5.78 &  14.09          &  4 F$_{\rm rot}$ \\
F7     &   8.04 &  12.59           & \\
F8     &    16.21  & 8.23           &11 F$_{\rm rot}$\\
F9     &    5.43 & 9.05         &3 F$_{\rm rot}$+1\\  
F10   &         10.70  &  8.92                  &9 F$_{\rm rot}$-1\\\hline
\end{tabular}
\end{table}

\begin{figure*}
  \begin{center}
 \includegraphics[width=18cm]{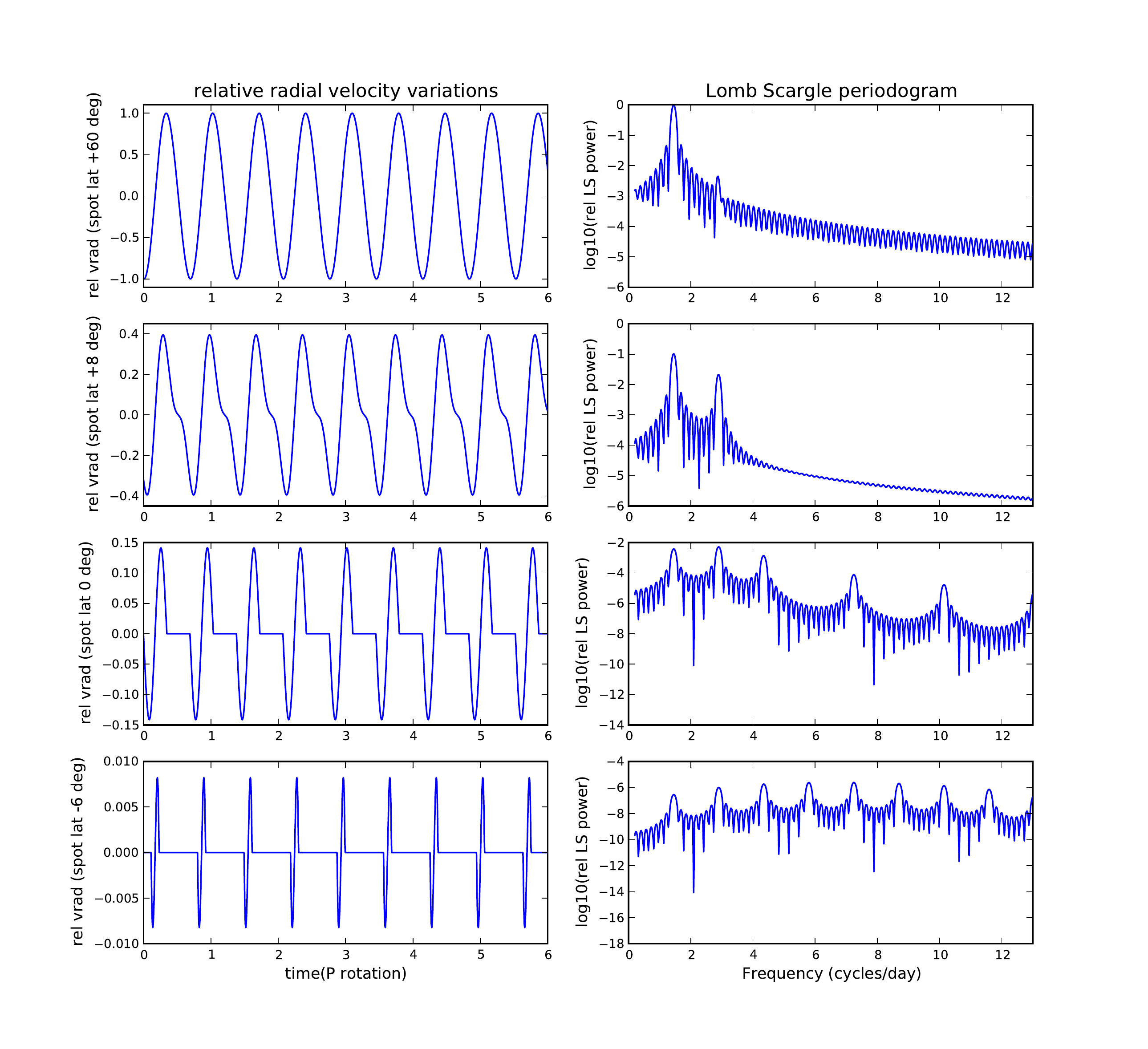}
   \caption{Simulation of the relative radial velocity variations induced by a spot (of same size) located at latitude (figures left, top to bottom): +60$^{\circ}$, +8$^{\circ}$, 0$^{\circ}$ and -6$^{\circ}$ for Vega (inclination angle i= 7$^{\circ}$ (the star is seen almost pole-on). Figures right, top to bottom show the respective Lomb Scargle periodograms. The spot position at latitude +60$^{\circ}$ is taken as the reference with respect to the radial velocity and Lomb Scargle power spectrum amplitude.}
   \label{fig:synth}
  \end{center}
   \end{figure*}

A possible coincidence, but nevertheless striking, is the fact that in the vspan periodogram not all harmonics are equally seen. F\,$_{\rm rot}$, 2\,F$_{\rm rot}$, 3\,F$_{\rm rot}$, 5\,F$_{\rm rot}$ and 7\,F$_{\rm rot}$ are seen, while 4\,F$_{\rm rot}$, 6\,F$_{\rm rot}$ and 8\,F$_{\rm rot}$ are absent. If vspan represents a cleaned radial velocity measurement, concentrating only on the activity within the line and not on its global shift, a simple simulation as shown in Fig.  \ref{fig:synth} can provide some explanation. Adopting the inclination of Vega's rotation axis with respect to the line of sight (i = 7$^\circ$), i.e. an star seen almost pole-on, we simulated using the basic trigonometric equations  the impact of a spot on the radial velocity, depending on its latitude. We normalized radial velocity and power in the Lomb Scargle periodogram for the spot at +60$^\circ$, and omitted any consideration on limb and gravitational darkening, true spot contrast and extension. The model is purely trigonometrical but shows nicely that
spots located close to the pole are seen permanently and provide an almost purely sinusoidal variation, yielding a single peak in the periodogram. Once the spot is at significantly lower latitudes a first harmonic appears. A spot location exactly at the equator will yield, due to the absence of spot structure during half the rotation period, the canceling of the 4\,F$_{\rm rot}$, 6\,F$_{\rm rot}$, 8\,F$_{\rm rot}$ harmonics. 
Does this provide an indication on the location of potential surface spots close to the equator? At even lower latitudes the spot is only seen for a very short time during the rotation period, yielding therefore all harmonics without canceling of any harmonics.

% FOLLOWING FIGURE NECESSARY? Shows impact of spot position on overall radial velocity variation...equations

%     montrer que si tache proche de 6 degre ou 7, alors sinc Prot/2 introduit Nu_rot (premiere harmonique), 2 nu_rot, 3 nu_rot, PAS 4 nu_rot, 5nu_rot, PAS 6 nu_rot, 7 nu_rot
%	(rajouter Herve Carfantan au papier?), ecrire les equations pour blobs equatoriaux (annulation de 4nu_rot...)? 

%     petit model (+ equation), montrer 2 taches (1 a +60, une a 6deg) ont un rapport d'amplitude trs different (celle a 5 domine tout). 

In a next step, we performed a freqeuency analysis of the variation of the two different radial velocity estimators. Fig. \ref{fig:ls_spec_vrad_bis} shows the result of a Lomb Scargle periodicity analysis of the variation of both radial velocity estimators with time. The corresponding frequencies are listed in Tables \ref{table:freqbis} and \ref{table:freqvmean}.

 \begin{figure*}
 \begin{center}
 \includegraphics[width=16cm]{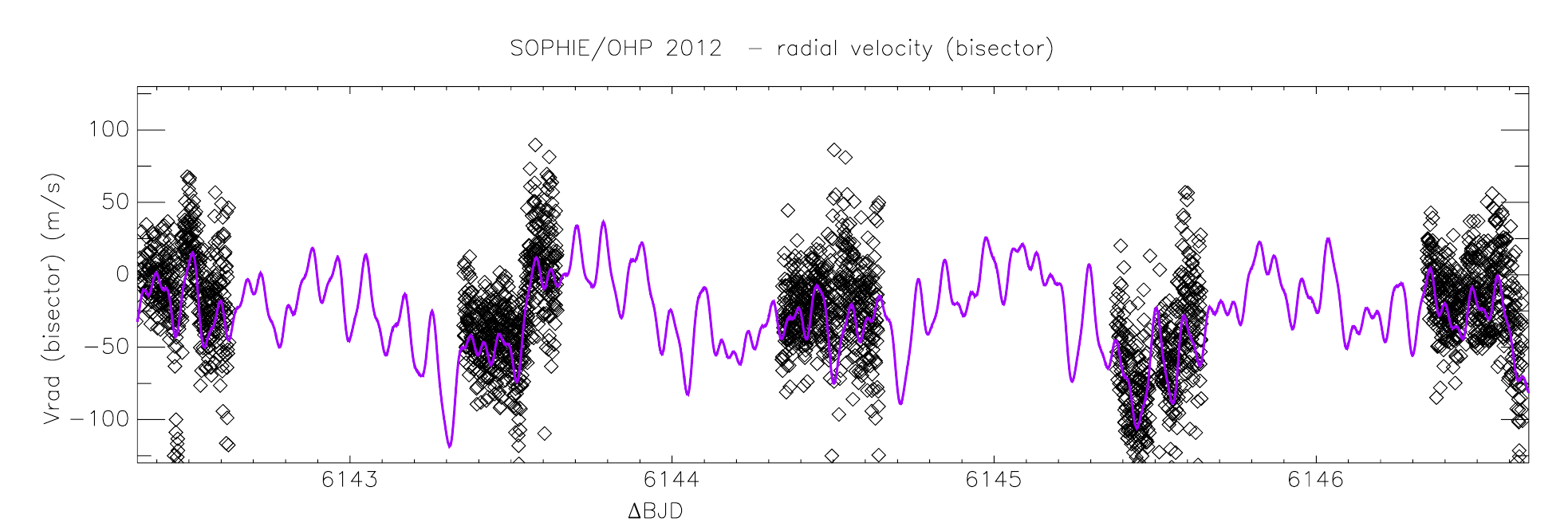}\\
 \includegraphics[width=16cm]{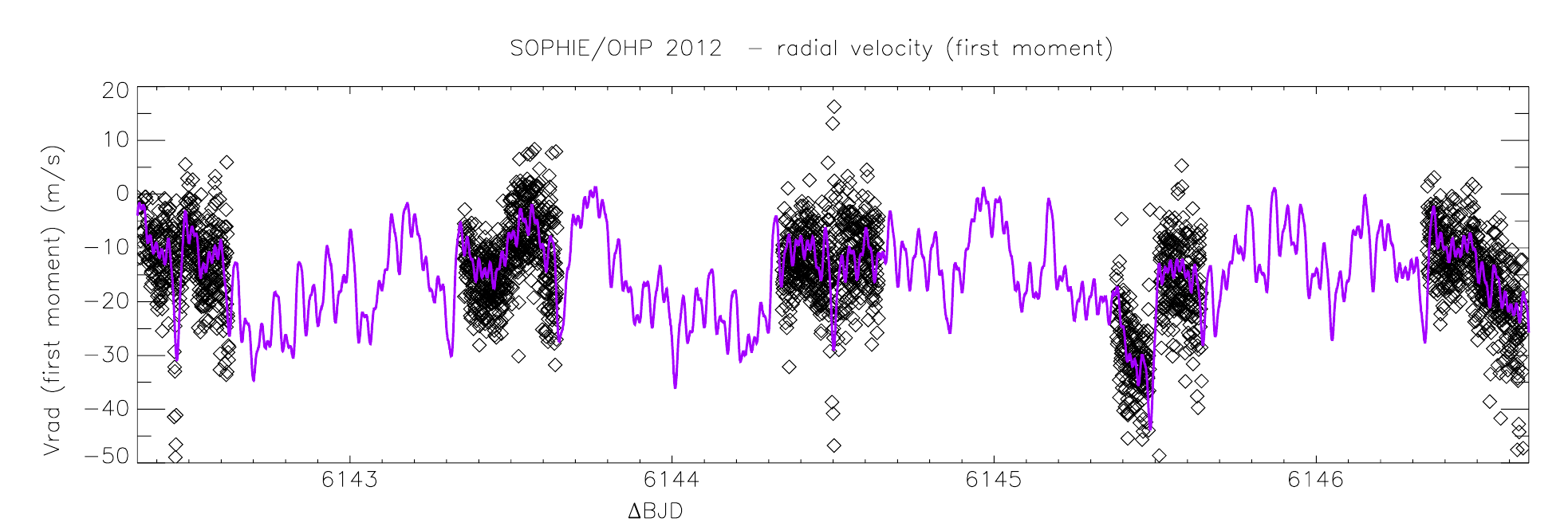}
   \caption{Radial velocity variations of Vega (SOPHIE/OHP) 2012 measured by the median of a lower bisector range (top) and first moment (bottom). Superimposed
(continuous line) is the result of the corresponding frequency analysis as listed  in Tables \ref{table:freqbis} and \ref{table:freqvmean}, respectively.
Time is expressed in BJD = 2450000 + $\Delta$\,BJD. Both RV determinations follow the same trend without being identical.} 
   \label{fig:time_vrad}
    \end{center}
 \end{figure*}

 \begin{figure*}
 \begin{center}
 \includegraphics[width=13cm]{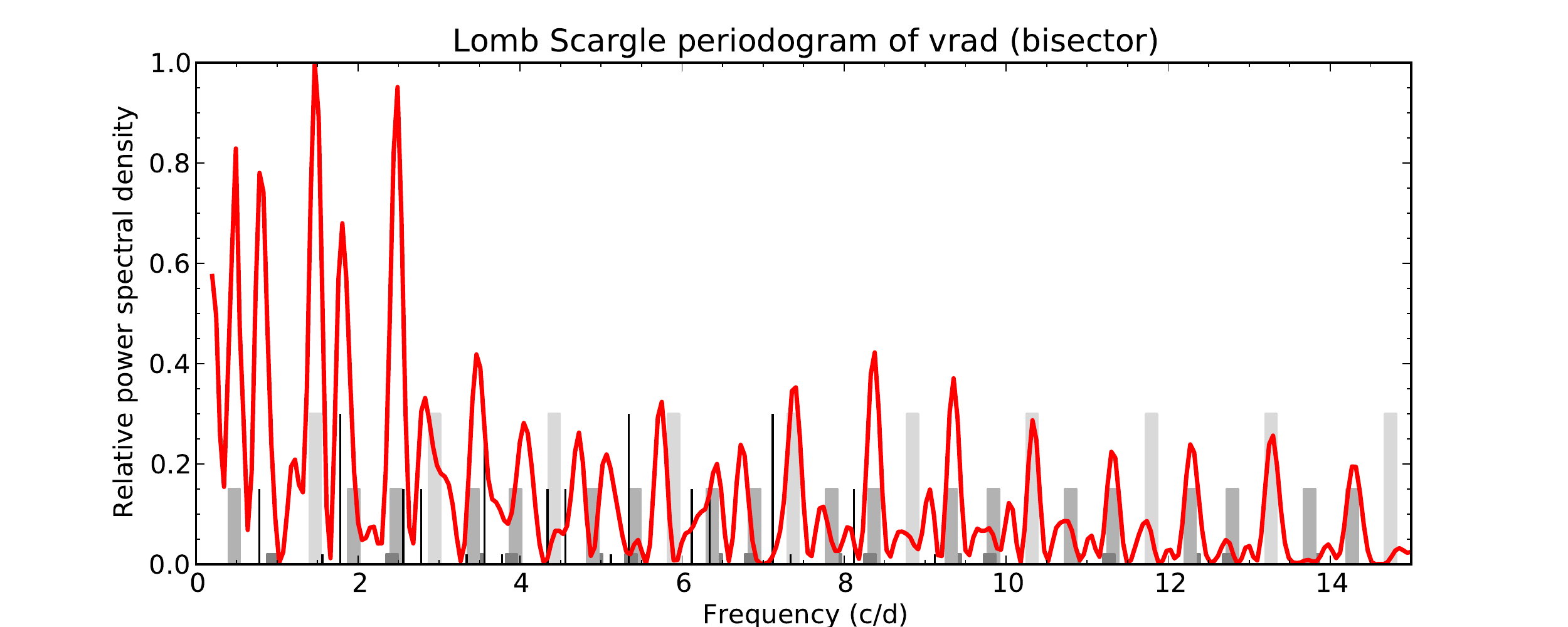}\\
 \includegraphics[width=13cm]{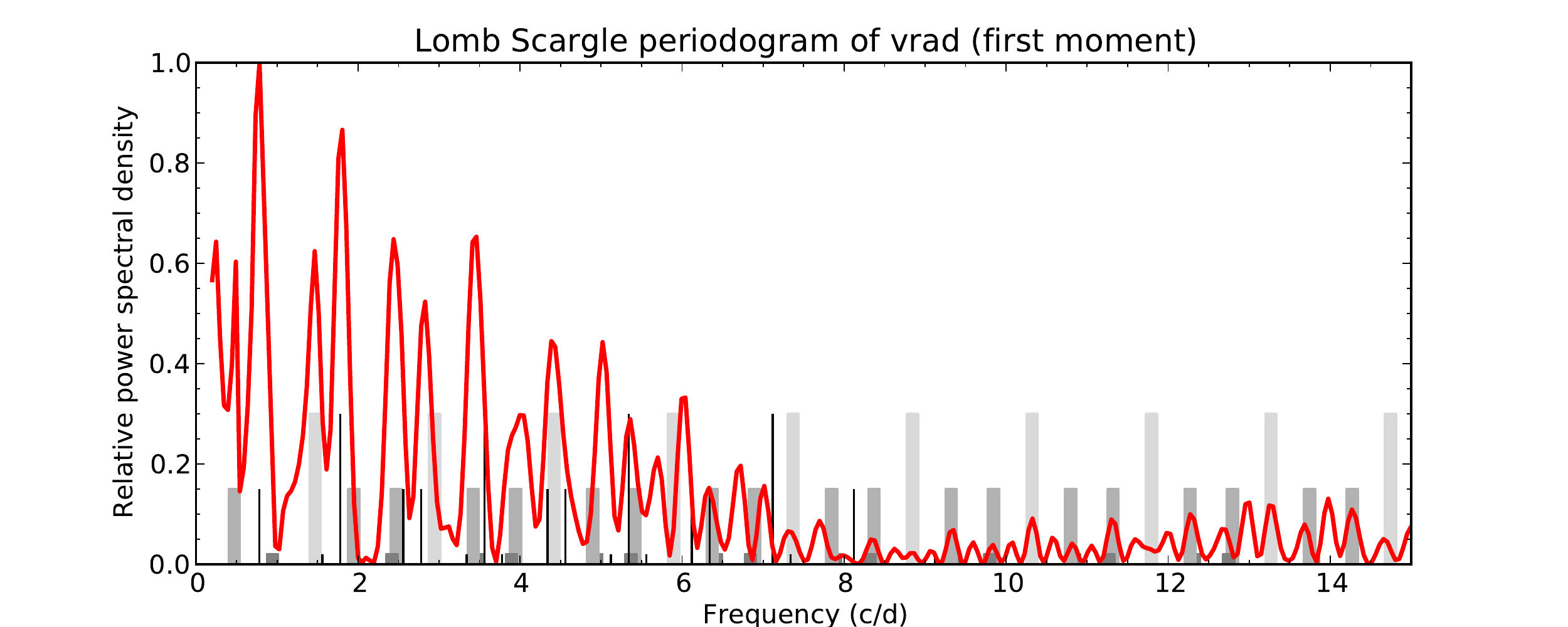}
   \caption{Lomb Scargle periodogram of the radial velocity evolution during the run, measured by (i) the lower part of the bisector (top) or (ii) the first moment of the profile (bottom).
  Definition of vertical bars and lines are identical to Fig. \ref{fig:ls_spec_vspan}. } 
   \label{fig:ls_spec_vrad_bis}
    \end{center}
 \end{figure*}
 
\begin{figure}
 \begin{center}
 \includegraphics[width=8cm]{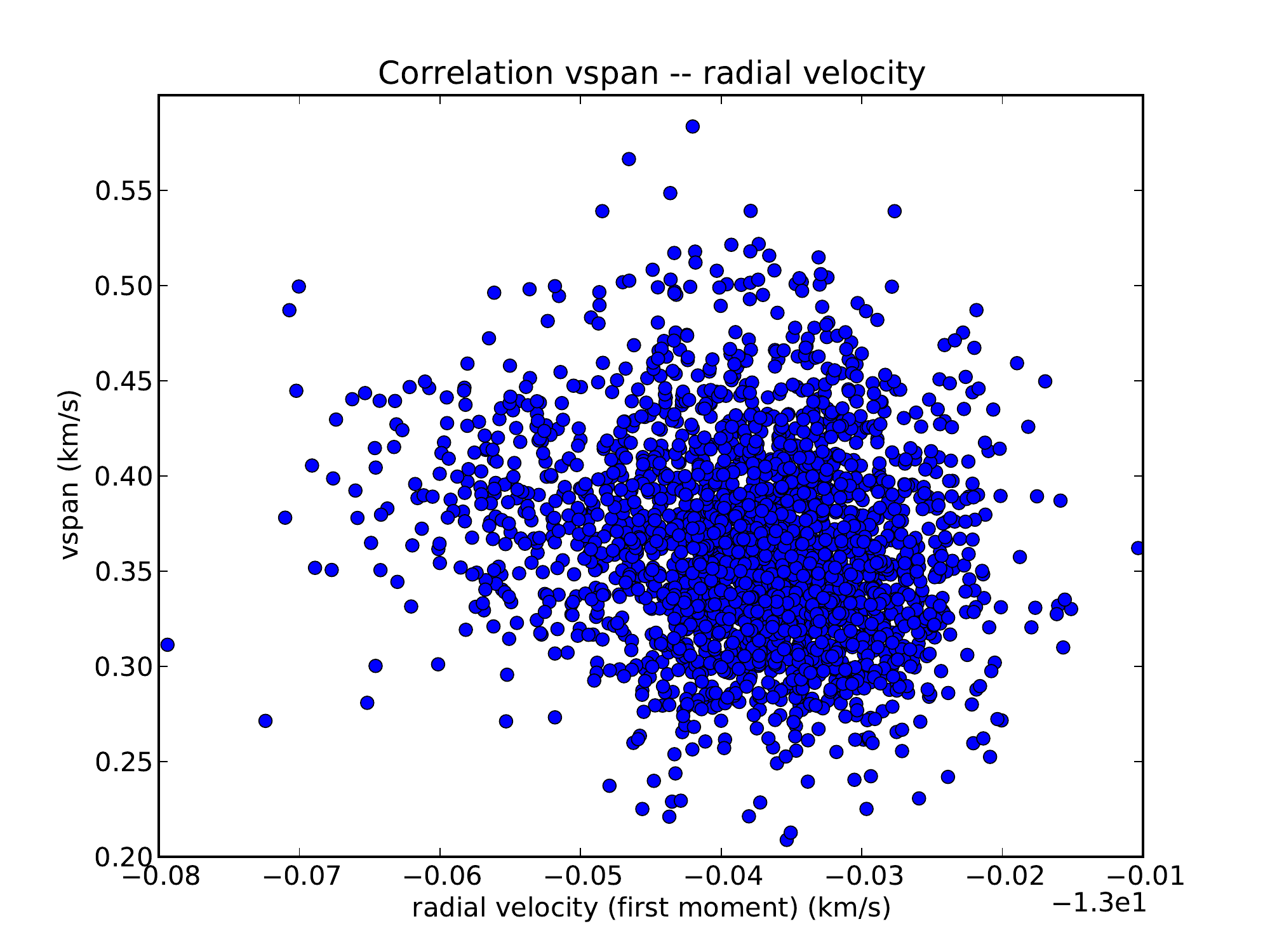}
   \caption{Anticorrelation of vspan versus vrad (first moment).} 
   \label{fig:vrad_mean_vspan}
    \end{center}
 \end{figure}

\begin{figure}
 \begin{center}
 \includegraphics[width=8cm]{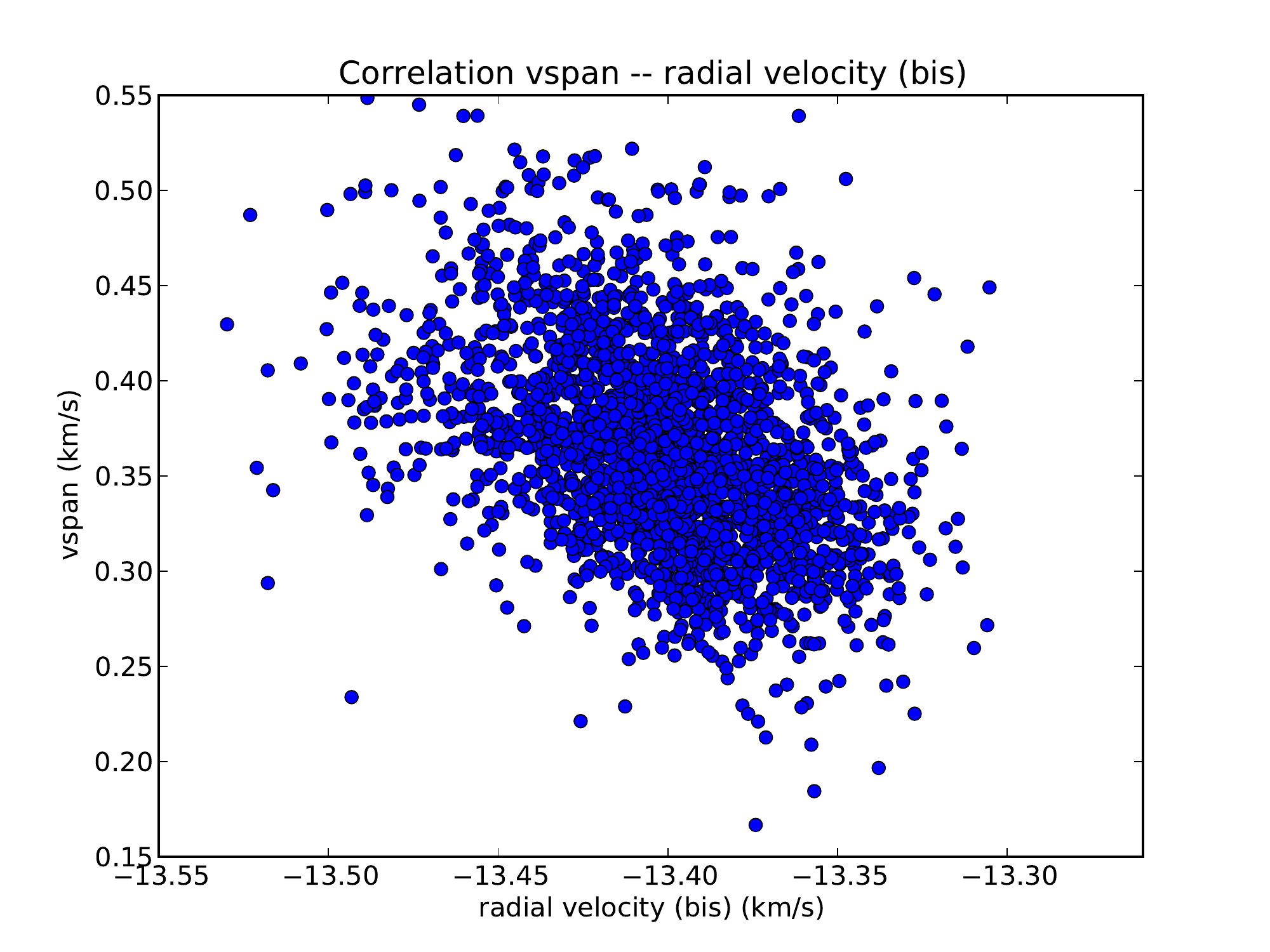}
   \caption{Anticorrelation of vspan versus vrad (bisector).} 
   \label{fig:vrad_bis_vspan}
    \end{center}
 \end{figure}

As can be seen in Fig. \ref{fig:vrad_mean_vspan} vspan is slightly anticorrelated to the radial velocity (first moment), which reinforces the idea of a starspotted surface. The anticorrelation is more or less pronounced mainly depending on the upper and lower bisector intervals chosen, more precisely, an upper bisector range moving towards higher values generates a stronger anticorrelation in our case. Several tests with different bisector ranges have been performed, also applying the traditional boundary values by \cite{queloz2001}. They yield similar results, and in all cases the strong presence of the rotational frequency is obvious. The method applied for radial velocity determination also reveals some impact on the level of anticorrelation as shown in  Fig. \ref{fig:vrad_bis_vspan}. Only a very weak effect can be seen as a function of the line selection minimum threshold, a higher threshold yielding a slightly more pronounced anticorrelation.

 During our analysis of the data set we did computations with different line masks, and some masks yielded a much more obvious anticorrelation. Since we concentrate on one specific line mask in this article we prefer showing the corresponding figure. 

The rotation period is again well represented in both data sets (F1b, F4c), and many harmonics and respective aliases of the window function are present. Amplitudes are different, which is due to different estimators. In the frequency analysis of the radial velocity (first moment) SigSpec extracts as highest amplitude frequency 0.77 \cd. We concluded however that Sigspec selected the -1d alias of F1c, which should be F1c = 1.77 \cd (and probably F2b = 1.89 \cd, despite the rather large difference in frequency). We compared two complete frequency grids (fundamental frequency at 0.77 \cd  or 1.77 \,cd, plus harmonics and window function aliases) with the observed periodogram and concluded based on the frequency distribution that the true frequency must be F1c = 1.77\cd, the grid for 0.77\, is much narrower and provides a significantly less good fit. The discussion of a potential origin of this frequency, which is not detected in vspan, is discussed in section \ref{osc}. 

 %PERIOD04
 
\begin{table}[!ht]
\caption{Frequencies and amplitudes of vrad (bisector) variations measured on the LSD-profiles (median value of lower part 0.15-0.3).}
\label{table:freqbis}
\centering
\begin{tabular}{|c|c|c|c|} \hline
ID      & freq.    & $A $ & Comment \\
        & \cd\,      & \ms &\\ \hline
F1b &    1.44 & 20.20 & F$_{\rm rot}$\\
F2b -1  &    0.89 & 22.79 & F2b = 1.89\\
F3b &    9.34 &  13.91 & 7F$_{\rm rot}$ - 1? F1$_{2008}$?\\
F4b &    13.32 & 9.52 & F2$_{2009}$\\
F5b &    4.15 &  9.08 & \\
F6b &    5.71&  11.46 & \\
F7b &    10.69 &  8.74  & 8F$_{\rm rot}$ - 1? F1$_{2009}$ -2 ? F2$_{2010}$ ?\\
F8b &    24.49 &  5.57  & 16F$_{\rm rot}$ + 1?\\
F9b &    9.94 &  6.17  &\\  \hline
\end{tabular}
\end{table}

\begin{table}[!ht]
\caption{Frequencies and amplitudes of vrad (first moment) variations measured on the LSD-profiles. Additional frequencies with low amplitude ($< 2.0$ \ms) have been seen at 17.54, 20.15, 25.11,  30.44 and 54.92 \cd.}
\label{table:freqvmean}
\centering
\begin{tabular}{|c|c|c|c|c|c|} \hline
ID      & freq.    &   $A$       & Comment\\
       	 & \cd\,     &   \ms        &          \\ \hline
F1c -1 &   0.77 & 6.16 	   &  F1c = 1.77 \\
F2c     &    3.34 & 4.32         &  3F$_{\rm rot}$ - 1? F2$_{2008}$ -2 ? F1$_{2010}$ -2 ? \\
F3c     &    4.97  &  3.78        &  4F$_{\rm rot}$ - 1?\\
F4c     &     1.48 & 4.01         & F$_{\rm rot}$ \\ 
F5c     &    24.47  &  2.17        &  16F$_{\rm rot}$ + 1?\\
F6c     &    14.24  &  2.42        & F2$_{2009}$ +1 ? \\
F7c     &    8.29  &  2.12        &  5F$_{\rm rot}$ + 1?\\
F8c     &      5.53 &  2.30        &  \\
F9c     &    11.78  &  2.61        & F1$_{2009}$ -1 ? F2$_{2010}$ +1 ?\\
F10c     &    14.67  &  2.68        &  10F$_{\rm rot}$\\ \hline
\end{tabular}
\end{table}

\subsection{Spot-like trails in dynamic spectra}
\label{spot}

We propose to investigate further the nature of photospheric features responsible for the 0.678\,d variability showing up in the time-series of radial velocities and velocity spans, and gather further evidence of the rotational origin of this periodical signal. 

\begin{figure}
 \includegraphics[width=7cm]{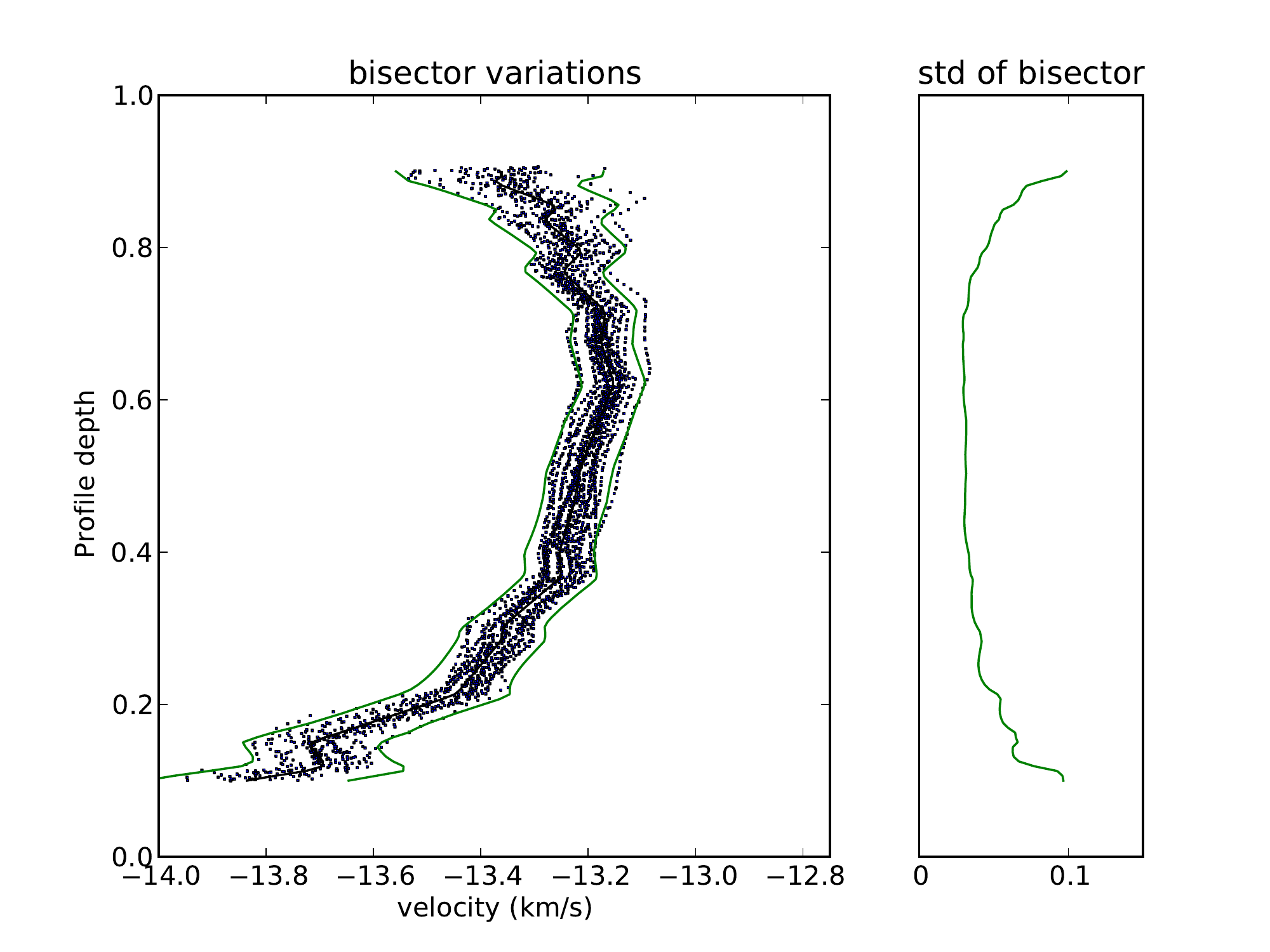}
   \caption{The distribution of bisectors in the total dataset is represented in the left figure. The red lines enclose 95\% credibility intervals for each depth. The right figure shows the standard deviation of the bisector variations as a function of renormalized profile depth. Note the enhanced activity of the bisector variations around the bottom of the profile, but also near the continuum. }
   \label{fig:bisectortime}
 \end{figure}

\begin{figure}
 \includegraphics[width=8cm, angle=-90]{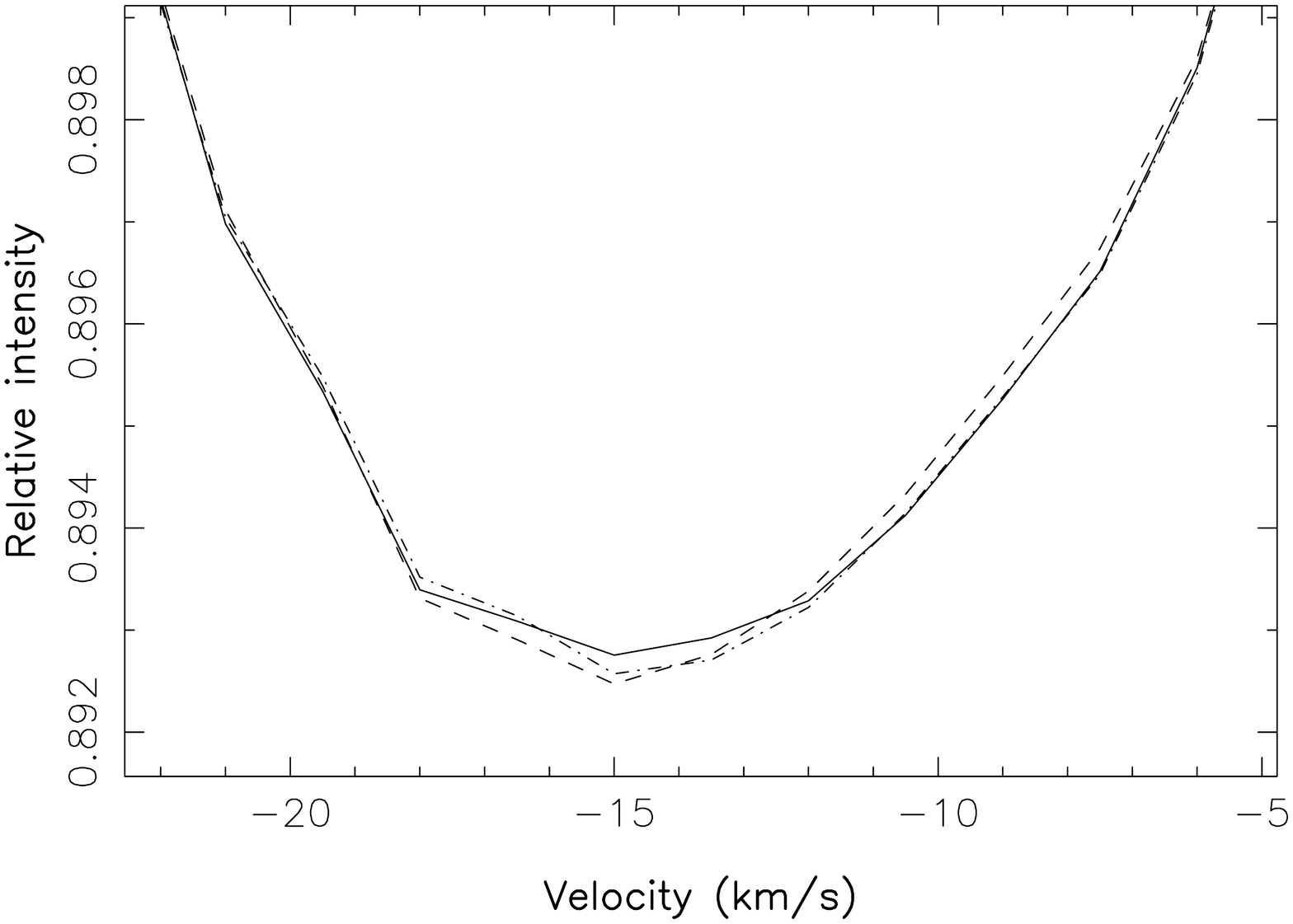}
   \caption{Variation of the LSD profile, averaged hour per hour, on the first three hours of the first night of the run (Aug. 2$^{\rm nd} \,2012)$. Travelling bumps can easily be seen. Continuous line: first hour (20-21:00 UT), dashed line: second hour (21-22:00 UT), dot-dashed line: third hour (22-23:00 UT).}
   \label{fig:variation}
 \end{figure}

Fig. \ref{fig:bisectortime} shows a D-shaped bisector, typical for early type stars \citep{gray2010}. As can be seen in this figure, enhanced "bisector activity" (as indicated by the spread in the left part of the figure, but also on the standard deviation as represented on the right side) of the bisector variations is seen around the bottom of the profile, but also near continuum. The variability of the bisector in the lower part is easily understood if surface structures cross the projected stellar disk during rotation. An easy explanation for the upper variations can not be provided at this stage. Very tiny LSD profile variations were seen in a thorough inspection (Fig. \ref{fig:variation}).

\begin{figure*}
 \begin{center}
 \includegraphics[width=16cm]{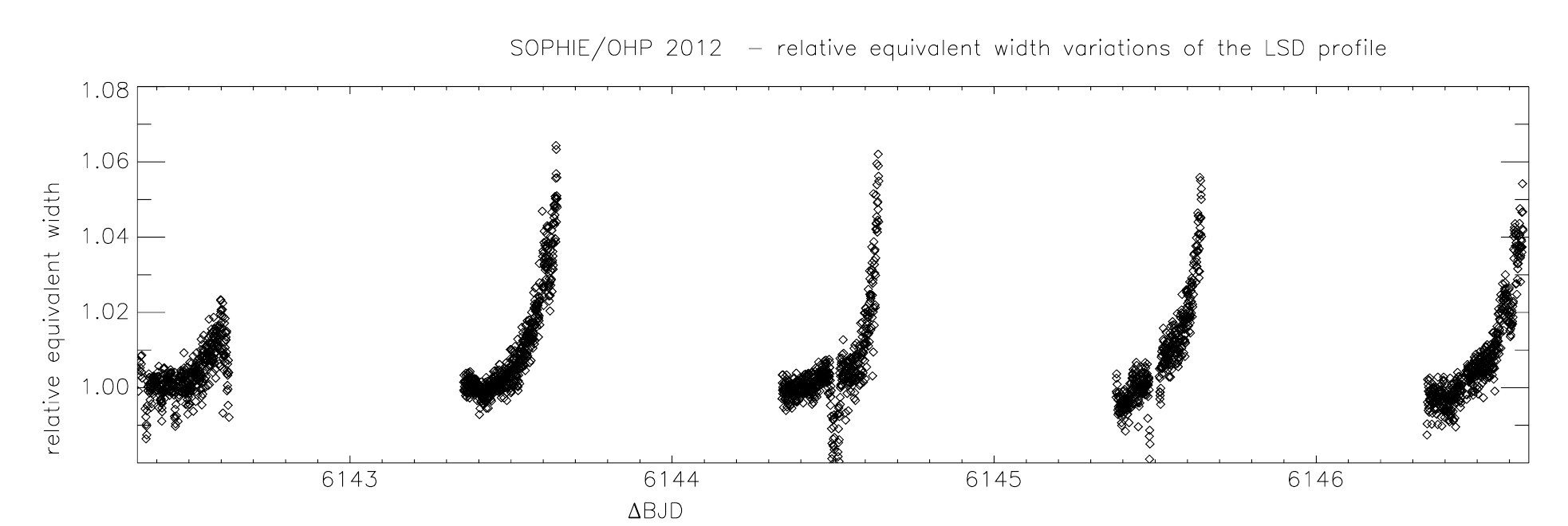}\\
   \caption{Nightly relative equivalent width variations of the LSD profiles of this run.
Time is expressed in BJD = 2450000 + $\Delta$\,BJD.} 
   \label{fig:eqwidth}
    \end{center}
 \end{figure*}

To highlight the tiny spectral features responsible for the periodic signal, we first cleaned up the data set by getting rid of LSD
equivalent photospheric profiles affected by the lowest S/N, therefore discarding about 1.4\% of all available data (files with less than 95\% of the mean S/N were rejected). 
As a next step, we corrected the LSD profiles from changes of their EW (equivalent width), observed as a systematic increase of the EW during each observing night. The daily
repeatability of this effect suggests an effect linked to the airmass of observations (Fig. \ref{fig:eqwidth}).

As this smooth evolution is observed to correlate with the S/N of LSD profiles, we can remove most of the systematic trend through the assumption that the equivalent width is a polynomial function of the S/N. For each night, a second order polynomial is adjusted to the equivalent width of LSD profiles as a function of the S/N, and the polynomial trend is then removed (ensuring that the average EW of the final distribution is the same as the initial one). We finally compute an averaged LSD profile for the entire run, that we subtract this mean from each LSD profile, to highlight fluctuations in the pseudo-line profiles.
It should be noted that the periodogram of the equivalent width variations (Fig. \ref{fig:ls_spec_eqwidth}) does not match any of the other periodograms (except the window function), and that it in particular doesn't show the rotational frequency or the 1.77\cd\, frequency.

  \begin{figure*}
\includegraphics[width=18cm,]{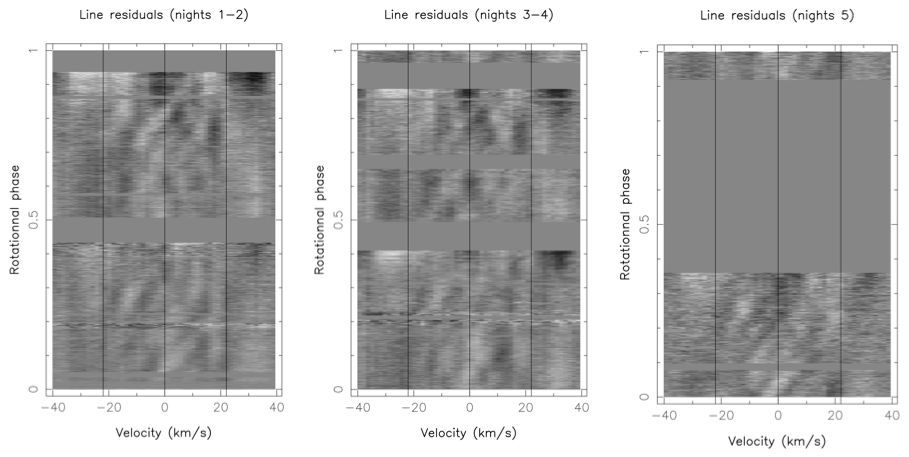}

   \caption{Residuals after nightly shift correction and subtraction of average profile are plotted as a function of velocity and time modulo period, where the period is fixed at the formally estimated value of P = 0.678\,d. The coherent structures along the diagonals in this plot are evidence of activity zones (in emission or absorption) moving together with stellar rotation and crossing therefore the line profile. The good matching of starspot features across several nights is a strong indicator for a structured surface on Vega.}
   \label{fig:phase2D}
 \end{figure*}

 \begin{figure}
 \begin{center}
 \includegraphics[width=9.5cm]{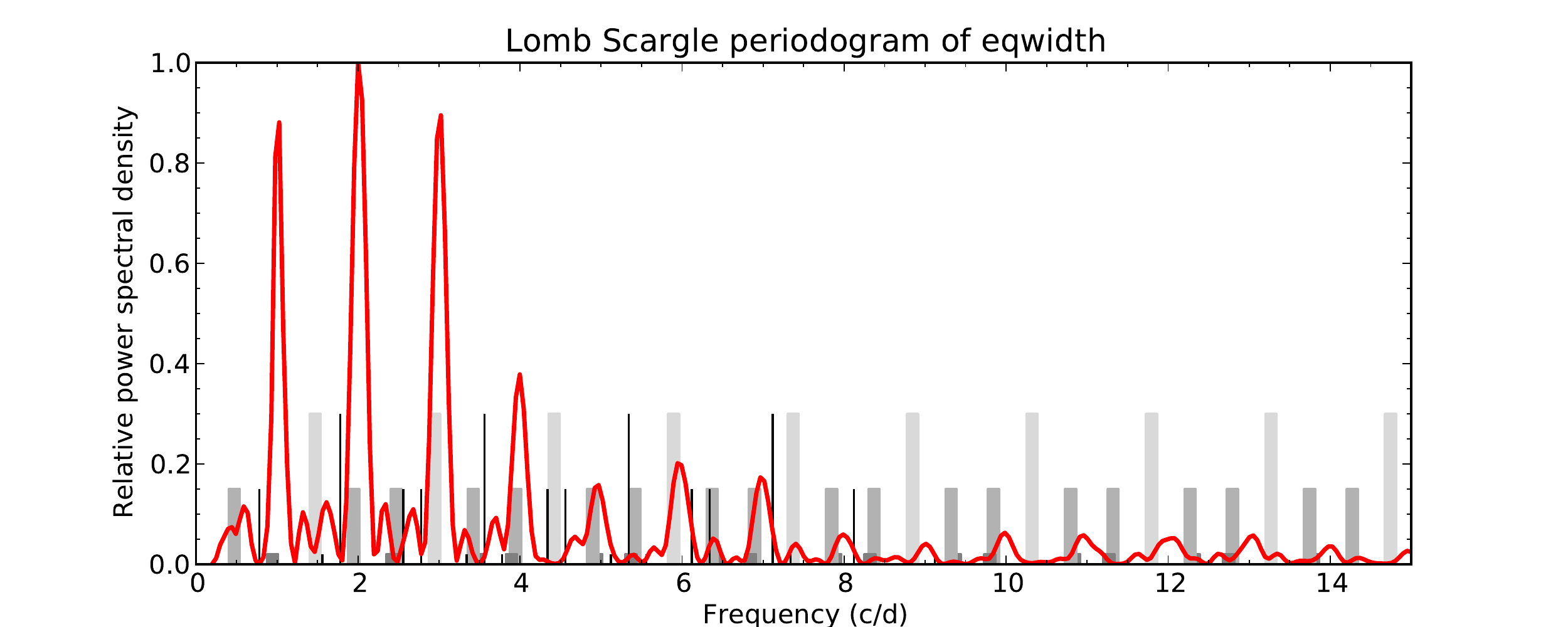}
   \caption{Lomb Scargle periodogram of the equivalent width variation during the run.
  Definition of vertical bars and lines are identical to Fig. \ref{fig:ls_spec_vspan}. } 
   \label{fig:ls_spec_eqwidth}
    \end{center}
 \end{figure}

The resulting line residuals are plotted in Fig. \ref{fig:phase2D}, after attributing a phase to each observation according to the 0.678\,d period, and choosing BJD=2456142.3308 (first stellar spectrum of the run) as the phase reference. 
The dynamic spectra display a number of bright and dark trail-like features first showing up in the blue wing of the line profile, and progressively shifting towards the red wing, in excellent agreement with the typical
spectral signature of brightness inhomogeneities carried across the visible stellar hemisphere during stellar rotation, as routinely observed in active solar-type stars \citep{2002MNRAS.330..699C}. These brightness inhomogeneities are sometimes referred to as "spots" and, if more extended, "plages", in the rest of the present article, regardless of their physical origin and brightness ratio relative to the quiet photosphere (darker or brighter regions). The amplitude of these subtle bumps and dips is of the order of $5 \times 10^{-4}$ -  $10^{-3}$ of the continuum. The 2D phase plots showing the transient spots are conclusive for all line-lists corresponding to different rejection thresholds. However some artefacts are present if the chosen threshold is too small  (a threshold of 0.1 or 0.2 
includes in the line-lists flat-bottomed profiles of weak lines which mix up with the mainly rotational profile of the stronger lines).

The resolution of the spectrograph is R = 75000, therefore the resolved element corresponds to 4 km/s in velocity. The typical width of each individual feature on the 2D phase plots of 
 Fig. \ref{fig:phase2D} are roughly $1/5^{\rm th}$ of the $vsini$ or 4.4km/s (see for instance trail at phase 0 and velocity 0).  The trail width are therefore barely resolved and we cannot make a conclusion on the minimum spot/plage size on the star's surface. Since the trail width are close to the resolution of the instrument, one could conclude however that their size on the stellar surface should be significantly smaller than the angular width 
corresponding to the trail size broadened by resolution, which is  $1/10^{\rm th}$ of $180^{\circ}$. It is therefore likely that sizes of the plages do not exceed $5^{\circ}$ in angular extension. The main trails are repeatedly observed during different observing nights, demonstrating that they do follow the 0.678\,d period and possess a lifetime of at least a few days. For example, the trail starting around 0 velocity at phase 0.1 can be seen every night at the same phase. The trail pattern (dark-bright-dark) close to -17 \kms reappears each night at phase 0.25. A backtraveling trail can be seen on nights 1-2 and 3-4 (night 5 did not cover this phase) starting at phase 0.5 and at +10 \kms  and reaching -10 \kms at phase 0.8. Many more examples can be found be comparing the three plots of Fig. \ref{fig:phase2D}. 
Note that a residual of EW variations is still visible in the dynamic spectra, at a level roughly similar to the spot-like trails (and is best observed as a systematically darker line core at the end
of each night).

If the trail amplitude is barely above noise level in the blue-red transit, the associated red-blue transit (visible in
principle for spots that are not eclipsed during stellar rotation thanks to their high latitude) are even fainter, owing to the less favorable projection factor and limb darkening during the return transit.
Still, a close look to the left and middle panel of Fig. \ref{fig:phase2D} reveals clearly back-traveling spot signatures.

In addition, a close inspection of individual trails further shows that the trail inclination in the dynamic spectrum is not identical for all spot-like features. 
As an illustration, the bright trail crossing each night the line center at phase $\approx$0.2 looks steeper than the other bright trail crossing line center 
at phase $\approx$0.05. The most natural interpretation of the different trail inclinations is a difference in the stellar latitude of 
brightness features.

% simul de trainee que bump laisse dans la moving_peak figure en fct de theta etc.

\begin{figure}
  \includegraphics[width=9cm]{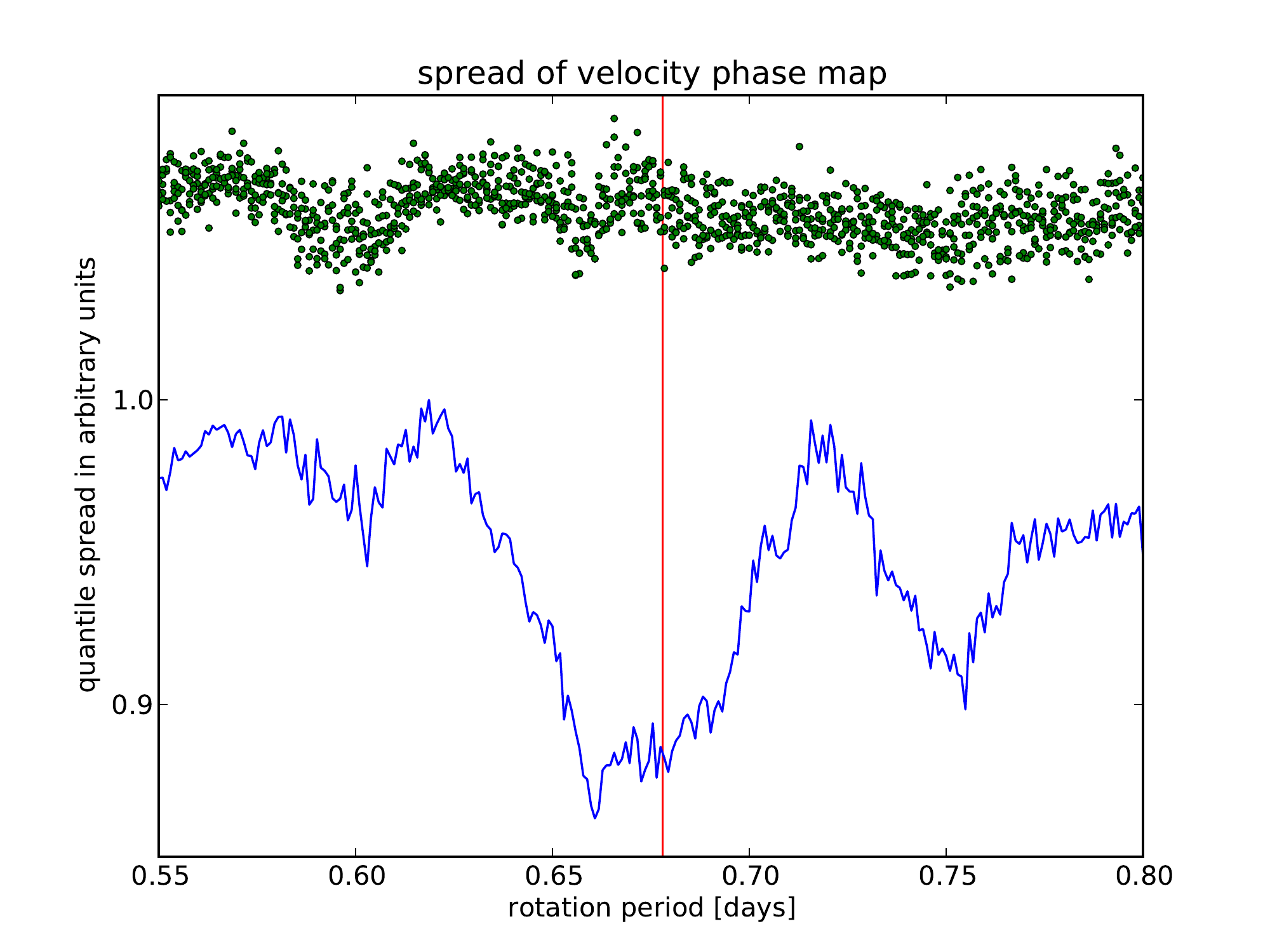}
   \caption{Optimal coherence is searched for a time remapping of the data set presented in Fig. \ref{fig:phase2D} around the pre-estimated rotational period. Optimality is based on the spread of the cloud in three dimensions (binned data in time modulo Prot, intensity, velocity). It can be seen that Prot = 0.678 d corresponds to the best matching period (blue line). To assess significance  we used the same data set with randomly permutated time data set (green dots).    }
   \label{fig:spreadphasemap.pdf}
 \end{figure}

Fig.  \ref{fig:spreadphasemap.pdf} reveals the results of a systematic attempt for optimizing the pre-estimated rotational period of P =  0.678\,d.
In practical, we scanned all periods in a range from 0.55 to 0.80\,d, around the period of rotation (as determined by \cite{alina2012} and this paper). After rephasing all spectra
with a given period, we calculated the average standard deviation within a binned subgrid in the velocity/phase space. The lowest consolidated standard deviation occurs when the rephasing period is optimized, i.e. when the residual intensities tend to be similar within the boxes of the subgrid. To assess significance we did the same analysis on a randomly shuffled data set (associating randomly a time value of the data set to a spectrum); the dotted cloud shows no minimum at all around the stellar rotation period and therefore clearly indicates that the observed minimum of our rephased data set is not due to random fluctuations of noisy data.

From this series of simple observations, we can draw two conclusions: (a) the 0.678\,d period can safely be interpreted as the stellar rotation period 
and (b) the spectral rotational modulation is produced by a complex pattern of spots peppering stellar surface. It is tempting to link these intricate brightness
inhomogeneities to the complex magnetic field topology previously reported by \cite{petit2010}.

The earlier data sets from 2008, 2009 and 2010 (\cite{boehm2012} ) were acquired with poorly stabilized spectrographs, and data reduction could only to some extend correct for these errors. 
Low frequency information was therefore totally lost, which is the reason why we did not directly detect the frequency of stellar rotation in their radial velocity data. Encouraged by our current results, we will try to present, in a forthcoming work, the results of a direct search for rotationally modulated LSD profiles in these data sets, a similar approach as presented in this section.

\subsection{Stellar oscillations and a potential exoplanet signature?}
\label{osc}

In \cite{boehm2012} we announced the presence of higher frequencies in the radial velocity periodograms of Vega, and suggested the possible detection of corresponding stellar oscillations. We only detected higher frequencies 5.32 and 9.19\,\cd\, (A $\approx$ 6\ms) in 2008, 12.71 and 13.25\,\cd\, (A $\approx$ 8\ms) in  2009 and 5.42 and 10.82\,\cd\, (A $\approx$ 3-4\ms) in 2010. As mentioned in the last section, these data sets were acquired with instruments not optimized in the sense of radial velocity stabilization,  making any low frequency detection,  such as stellar rotation, impossible.

As can be seen in tables \ref{table:freqbis} and \ref{table:freqvmean} corresponding to our 2012 run, some energy in the periodograms is located in the higher frequency domain.  Possible identifications with frequencies found in the former data sets are indicated. Still, the dense frequency spacing of the rotational harmonic comb and its different window function aliases, together with propagated error bars, do almost always allow to find a possible identification with rotationally linked frequencies. 
This tells us that rotation might be in all data sets responsible for the observed higher frequencies too. However, it still does not exclude the presence of oscillations in all these data sets, it just prohibits any conclusion at this stage.

\begin{figure*}
 \begin{center}
 \includegraphics[width=18cm]{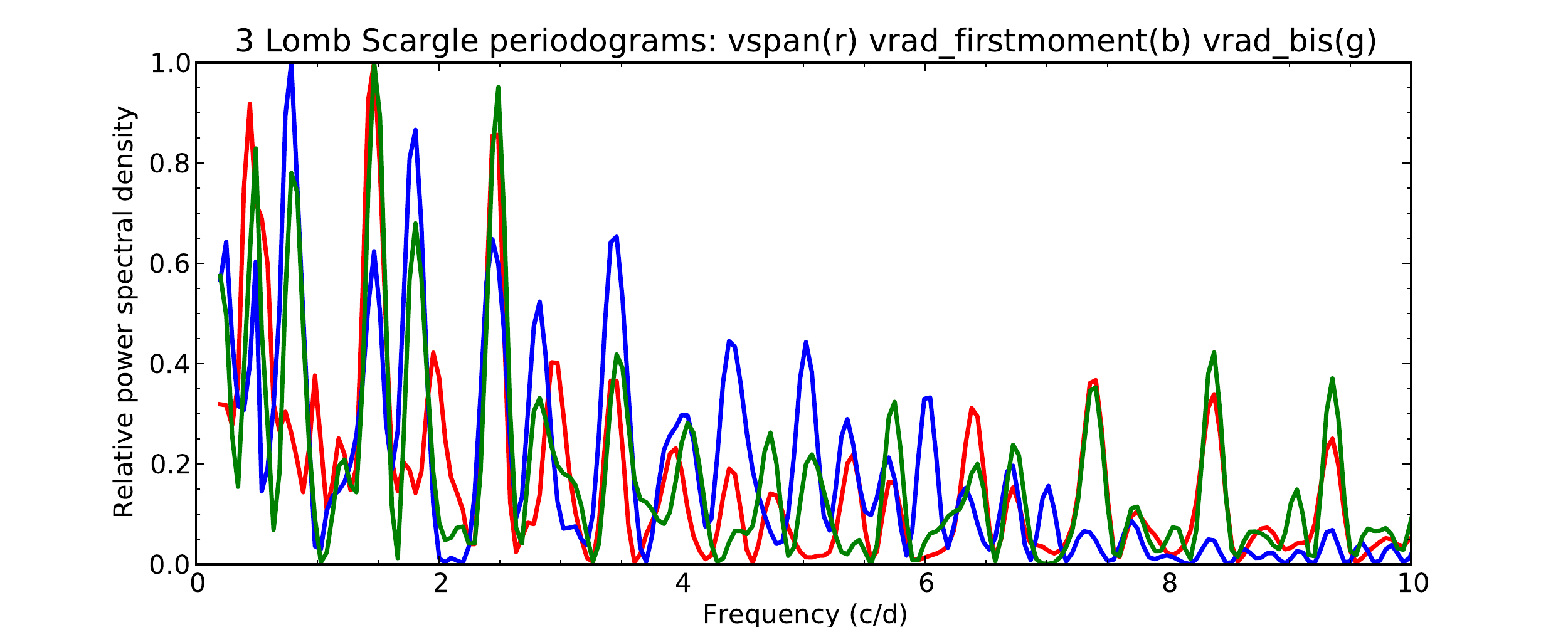}
   \caption{Lomb Scargle periodogram showing the frequencies present in vspan (red), vrad (first moment, blue) and vrad (bisector, green). It can clearly be seen that all three
   estimators show the rotation frequency, but only the two radial velocities show the 1.77\cd frequency. Color codes are online. } 
   \label{fig:ls_spec_all}
    \end{center}
 \end{figure*}

The most striking difference between the periodicity analysis of vspan and the two different radial velocity determinations is the fact that the strong frequency at 1.77 \cd \, (or 1.89 \cd)  only appears in the latter ones. 
(see Fig. \ref{fig:ls_spec_all}). For the sake of completeness we want to mention that these additional frequencies do not appear when selecting only the strongest lines (threshold of 0.7). In that case the number of lines drastically diminishes to only 144 lines and all periodograms become messy, without any clear structure.  For a threshold of 0.1 and 0.3 the number of  lines used for the LSD correspond to 1494 and 296, respectively, yielding a much higher S/N equivalent photospheric profile.
Vspan measures an asymmetry within the profile and is totally insensitive to global radial velocity shifts. Should we therefore understand F1c =  1.77 \cd  (or F2b) as the signature of a global, i.e. dynamical shift of the line profile? The presence of an exoplanet could be one possible source of such bulk RV variations. Activity induced radial velocity signature (starspots/ bumps traveling through the line profile) could not provide this type of global variations. In Fig. \ref{fig:ls_spec_vrad_bis} one can see that F1c is most likely in presence of two harmonics at 2,F1c and 3\,F1c. A low excentricity orbit could explain a low number of harmonics. 
Since amplitudes differ significantly between both radial velocity measurements, only very gross deductions can be tempted. Applying Kepler's laws and Vega's fundamental parameters, using the global shift measuring radial velocity determined by the first moment,  and supposing an exoplanet orbiting in the equatorial plane at a distance of 0.017 AU, a 0.34 M$_{\rm Jupiter}$ exoplanet could satisfy the observed parameters (stellar mass: 2.15\,M$_{\odot}$, amplitude v$_{\rm orb (Vega)}$sini $\approx$ 6 \ms \,derived from the first moment measurement, inclination angle of the system: 7$^{\circ}$ (pole-on), P = 0.56\,d (corresponding to F1c =  1.77 \cd), eccentricity e = 0). Using the values of vrad (bisector), ie. P = 0.53\,d and v$_{\rm orb (Vega)}$sini $\approx$ 22.9 \ms\, an exoplanet mass of 1.24 M$_{\rm Jupiter}$ at 0.0165 AU would satisfy the equations. In both case studies, the interesting result is the proximity of the frequency to the rotation frequency. It indicates that the orbital radius of such an exoplanet would correspond to only 1.36 (or 1.31) R$_{\rm Vega}$ (calculated for e = 0 and using the concordance model as published by \cite{monnier2012}), while the co-rotating radius /synchronous orbit corresponds to 1.5 R$_{\rm Vega}$. If we have at this stage no further indication supporting this potential presence of an exoplanet and knowing that until now only few exoplanets with such short orbital periods have been detected, attention should be given to recent results  by \cite{balona2014}, announcing the fact that approximately 19\% of A-type stars were potentially accompanied by a roughly Jupiter mass exoplanet in a synchronous orbit. The most interesting conclusion of our (still hypothetical) analysis is therefore that planetary material of a significant fraction of a Jupiter mass could be located close to the synchronous (or co-rotation) radius.  This result  seems to agree with the findings of \cite{balona2014}, and Vega could be one of this close-planet A-type stars. If confirmed, it would be very interesting to observe associated tidal effects of such a close in planet orbiting a "hot" star. Would these findings indicate that in A-type stars exoplanet migration stops at the corotating radius? Would that imply that dipolar magnetic fields dominate regions out to the corotation radius and suppress any further migration?

\section{Conclusion}
\label{disc}

The discovery of corotating structures at the surface of a non-chemically peculiar A-type star provides
new insights into the processes at work in the envelope of a typical intermediate-mass star.
Long-lived large scale chemical spots are already known to exist on Ap/Bp stars, but this class of magnetic and 
chemically peculiar stars only represents 5-10 \% of stars in this mass range and the origin of their spots is attributed to the atomic diffusion of chemical elements in an atmosphere
stabilized and structured by a strong 300 Gauss or higher large scale fields. 
Chemical anomalies associated with abundance  spots are also seen in HgMn stars, another class of chemically peculiar stars with late-B spectral types (e.g., \cite{makaganiuk2011, korhonen2013}). 
No evidence of magnetic fields has stood up to scrutiny, and upper limits as low as a few G exist for some stars \citep{kochukhov2013}. These stars show slow rotation rates for hot stars, a property which most likely helps in creating a stable atmosphere where atomic diffusion can operate. In addition to the rotation rate, there is a very significant difference between Vega and these late B-type stars, namely that the late B stars have undetectable micro turbulence, which indicate that they have stable atmospheres. In contrast, Vega has non-zero micro turbulence a property consistent with the abrupt change in micro turbulence velocity at roughly the A0-B9 boundary \citep{landstreet1998}.

%It is hard to see how patches could persist for a long time in such an atmosphere, although they might develop dynamically, like the Red Spot on Jupiter. Stability in the atmosphere means within observable optical depths. In a general way, below about 10000 K, the atmospheres of main sequence stars are unstable in the atmosphere and somewhat below due to high opacity in the H ionization zone. In such stars the value of the microturbulence parameter seems to be a pretty good proxy for convection in visible layers. Above about 11000 K the excess of the radiative gradient above the adiabatic one is very slight and it is not clear if there is significant convection; in any case the microturbulent broadening becomes undetectable even in very sharp line stars. So there is an important atmospheric change at roughly the A0-B9 boundary. All of these stars probably have substantial convection in the region where the ionization zone of He II -> He III provides strong opacity, and they may have rather weak convection just below the atmosphere where He I -> II ionization leads to strong opacity. They may also have the iron bump opacity at 10^5 K. So the main difference between A and B stars is the strength of convection actually IN the atmosphere (strong in A stars, weak or absent in late B stars); they all have some convection in deeper layers. 

In a non-chemically peculiar A-type star like Vega, the rotational modulation must be of different origin since with a very low magnetic field and a high rotation rate the atmosphere can not be stable enough to generate chemical anomalies through atomic diffusion. Nevertheless the magnetic field of Vega is still a natural explanation for corotating structures seen with Doppler Imaging.
In this case, the property of these structures should help determine the enigmatic nature of Vega's magnetic field.

Although the envelope of A-type stars is mainly radiative, small convective layers due to respectively the hydrogen, the first and the second helium
ionizations are present. This layers can in principle host a dynamo driven by the convective motions. 
Due to their small thickness and low density, the energy contained in these motions is limited. Moreover, close the surface, the convective turnover time can be
much smaller than the rotation rate, in which case the dynamo is inefficient. Nevertheless, \cite{cantiello2011} found that in the hotter O and B stars the
convective layer due to the opacity peak related with iron group elements can generate relatively strong magnetic field, assuming equipartition between kinetic and magnetic energy
and estimating the convective velocity from the mixing length model. In A-type stars, the convective layer induced by the second helium ionization \citep{weiss1999} might play a similar role.
However, whether such a dynamo can generate the observed $7$ Gauss nearly polar spot of Vega \citep{petit2014} and also 
account for the observed co-rotating structures remains to be verified. Furthermore,
convective dynamos are intrinsically variable with a spot lifetime of the order of the rotation period. Such a variability is not detected in our data. 

Other possible origins of Vega's magnetic field, discussed by \cite{lignieres2009} and  \cite{braithwaite2013}, rather involve fields generated in the early phase of the star life and their 
subsequent evolution in the radiative enveloppe. In this context also, a key feature that would help distinguish between models is the intrinsic time variation of the field. Given the low-amplitude of the Stokes V profile, we expect that, if present, this variability will be easier to detect through spectroscopic studies similar to the present one.

Another question raised by the present study concerns the sign and the amplitude of the luminosity contrast induced by the co-rotating structure.
For the weak magnetic field observed in Vega, bright rather than dark spots are expected because dark spot only occur when the field is strong enough to limit convective heat transport within the
spot.  As mentioned before, rotational modulations compatible with spots have been detected with Kepler's light curves in a large fraction of A-type stars  \citep{balona2011}.
If these modulations are the photometric counterpart of the present spectroscopic structures, this would strongly support the existence of a widespread Vega-like magnetism
and activity among A-type stars.

\begin{acknowledgements}   
The author wants to thank the staff of SOPHIE/OHP  for their efficient support during these challenging observing runs. He also acknowledges support from the french national program PNPS/INSU and the ANR project Imagine. M.R. acknowledges financial support from the FP7 project SPACEINN: Exploitation of Space Data for Innovative Helio- and Asteroseismology. GAW is supported by a Discovery Grant from the Natural Science and Engineering Research Council (NSERC) of Canada. 
\end{acknowledgements}

\bibliographystyle{aa}
\bibliography{vega_2.2_arxiv}

\end{document}